\documentclass[fleqn,usenatbib]{mnras}

\usepackage{newtxtext,newtxmath}

\usepackage[T1]{fontenc}

\DeclareRobustCommand{\VAN}[3]{#2}
\let\VANthebibliography\thebibliography
\def\thebibliography{\DeclareRobustCommand{\VAN}[3]{##3}\VANthebibliography}

\usepackage{graphicx}	
\usepackage{amsmath}	
\usepackage{textgreek}

\newcommand{\ldust}{\hbox{$L_{\rm dust}$}}
\newcommand{\mdust}{\hbox{$M_{\rm dust}$}}
\newcommand{\mstar}{\hbox{$M_\ast$}}
\newcommand{\msun}{\hbox{$M_\odot$}}
\newcommand{\agem}{\hbox{age$_M$}}

\newcommand{\eagle}{\hbox{\textsc{eagle}}}
\newcommand{\magphys}{\hbox{\textsc{magphys}}}
\newcommand{\skirt}{\hbox{\textsc{skirt}}}

\title[Testing the \magphys\ SED code]{How well does \magphys\ recover galaxy properties? A test using \eagle\ simulated star-forming galaxies}

\author[Jones, da Cunha \& Battisti]{
Zoe R. Jones$^{1,2}$\thanks{E-mail: zoe.rhiannon.jones@ut.ee}, Elisabete da Cunha$^{1,3}$\thanks{E-mail: elisabete.dacunha@uwa.edu.au}, and Andrew Battisti$^{1,3,4}$
\\
$^{1}$International Centre for Radio Astronomy Research (ICRAR), The University of Western Australia, M468, 35 Stirling Highway, Crawley, WA 6009, Australia\\
$^{2}$Tartu Observatory, University of Tartu, Observatooriumi 1, 61602 Tõravere, Estonia\\
$^{3}$ARC Center of Excellence for All Sky Astrophysics in 3 Dimensions (ASTRO 3D), Australia\\
$^{4}$Research School of Astronomy and Astrophysics, Australian National University, Cotter Road, Weston Creek, ACT 2611, Australia}

\date{Accepted 2026 March 20. Received 2026 March 20; in original form 2025 June 06}

\pubyear{\the\year{}}

\begin{document}
\label{firstpage}
\pagerange{\pageref{firstpage}--\pageref{lastpage}}
\maketitle

\begin{abstract}
Spectral energy distribution (SED) models are widely used to infer the physical properties of galaxies from multi-wavelength photometry, but their accuracy is difficult to assess because the true properties of observed galaxies are generally unknown. We address this by fitting synthetic SEDs of $\sim31,000$ star-forming galaxies drawn from the \eagle\ cosmological simulations, post-processed with the \skirt\ radiative transfer code, using the \magphys\ SED modelling framework. This provides a controlled testbed with known intrinsic parameters, enabling a direct assessment of model accuracy and the origin of systematic biases.
Under idealised conditions, fitting well-sampled ultraviolet-to-submillimetre SEDs at $z=0.1$, $z=2$, and $z=5$, \magphys\ recovers stellar mass, star formation rate, specific star formation rate, dust mass, and dust luminosity to within $\lesssim0.14$~dex, while mass-weighted stellar ages are not robustly constrained. We find that mismatches between the assumed star formation history (SFH) priors and the intrinsic SFHs of the simulated galaxies introduce systematic biases in stellar mass estimates, even when the fits provide good statistical agreement.
To assess performance under realistic survey conditions, we construct a WAVES-like mock sample using optical and near-infrared photometry with realistic uncertainties. In this case, stellar masses and star formation rates remain well constrained (systematic offsets $\lesssim0.1$~dex; scatters $\simeq0.07$ and $\simeq0.15$~dex, respectively), whereas dust properties degrade significantly without far-infrared data: dust luminosities show offsets of $\simeq0.30$~dex and scatters $\simeq0.25$~dex, and dust masses exhibit scatters $\simeq0.3$~dex.
We conclude that MAGPHYS is a reliable tool for recovering key galaxy properties from broad-band photometry, but that SFH assumptions and limited wavelength coverage introduce significant uncertainties, particularly for dust and stellar ages.

\end{abstract}

\begin{keywords}
galaxies: fundamental parameters -- galaxies: photometry
\end{keywords}



\section{Introduction} \label{sect: Introduction}

The multi-wavelength emission, or spectral energy distributions (SEDs), of galaxies encode much of their critical information, such as their stellar masses and star formation rates (see reviews by, e.g., \citealt{Walcher11,Conroy13,Madau14}). However, interpreting these parameters from observed galaxy emission alone is complex, due to numerous degeneracies in SED colours that arise from the complex combinations of emitting and attenuating sources (e.g., stellar age and dust attenuation in ultraviolet (UV) to optical colours). Nevertheless, various SED models have been developed over the past decade and a half that aim to robustly model the observed SEDs of galaxies at various redshifts, and constrain their physical properties (e.g., \citealt{magphys, Noll09, BEAGLE, BAGPIPES, PROSPECTOR, PROSPECT}; see recent review by \citealt{Pacifici23}). The vast majority of these SED models include a spectral synthesis model for the emission of stellar populations (e.g., \citealt{Bruzual03}), combined with some assumption on the stellar initial mass function (IMF) and star formation history (SFH), a dust attenuation prescription (e.g., \citealt{Calzetti94, Calzetti00, Charlot00}; see \citealt{Salim20} for a review), and an energy balance method to connect the UV, optical, and near-infrared stellar emission consistently to the mid- and far-infrared dust emission (e.g., \citealt{magphys}). SED models therefore rely heavily on simplifications and assumptions to model galaxy emission, with the goal of restricting the parameter space to save computing time while still being realistically applicable to large data samples. The exact computation of true-to-life galaxy SEDs would require more computationally-intensive radiative transfer models (e.g., \citealt{Baes10, Popescu11, Narayanan21}), which are at present not easily applicable to the extraction of physical parameters from large galaxy surveys.

A main challenge of frequently used SED models comes from determining how reliable their recovered parameters are. Several studies have compared some of the available SED models with each other to check for systematic differences between different codes (e.g., \citealt{Hunt19, Pacifici23}), while others have compared the more fundamental stellar population models, dust models, and photometry choices, and their effects on parameter recovery (e.g., \citealt{Maraston10, Siudek24, Tortorelli24}). However, this still leaves us with the question of how well SED models recover the true physical parameters of observed galaxies. That is, such studies sought to find which SED codes agree most with each other in order to predict their accuracies, rather than comparing them to a ground truth. Given the degeneracies in galaxy SEDs, and the relatively large number of free parameters in these models compared to the number of observables (usually, multi-wavelength broad-band fluxes), simply checking the goodness-of-fit to the observations is not sufficient to tell us whether a model fit is robust or not, and whether we can trust the recovered parameters. SED models try to overcome these degeneracies and account for uncertainties by adopting Bayesian frameworks that produce the full likelihood distributions of the physical parameters, and allow one to assess the impact of degeneracies and uncertainties. Still, the absolute calibration of the recovered parameters remains a challenge.

One way to test the outputs of SED models is to apply them to galaxies for which we know the `ground truth' using other methods. For example, \cite{Battisti19} developed the photometric redshift extension of the \magphys\ model \citep{magphys,daCunha15} and, when testing this code on galaxies with known spectroscopic redshifts from the COSMOS survey \citep{Scoville07, Weaver22}, they found a better agreement between the \magphys\ photometric redshifts and the spectroscopic redshifts when a 2175 \AA\ feature was included in the model dust attenuation curves, rather than using the featureless power-law model of \cite{Charlot00}. This proved to be an independent way of testing SED model assumptions on real galaxies. Another way to do this is to use mock galaxies, as in the work of \cite{Hayward15}, who used the \textsc{sunrise} Monte Carlo dust radiative transfer code \citep{Sunrise06, Sunrise10} to create synthetic SEDs of two \textsc{gadget-3} \citep{Gadget01, Gadget05} simulated galaxies for which the physical parameters are known: an isolated disk galaxy, and a merger of two disk galaxies. They then used the \magphys\ model to fit these synthetic SEDs and constrain the parameters of the simulated galaxies, and compared them to the `true' known parameters (see also \citealt{Dudze20} for a similar test in the context of recovering the physical parameters of sub-millimetre galaxies). \citet{Hahn25} conducted a similar study where they fit the synthetic SEDs of the \textsc{nihao+skirt} catalogue \citep{Wang15, Faucher23} with the \textsc{provabgs} SED code \citep{Hahn23, Kwon23}, finding the impact of assumed dust geometry on parameter recovery. \citet{Narayanan24} conducted a similar study at high-redshift using galaxies from the \textsc{simba} simulations \citep{Dave19}, processed by the \textsc{powerday} dust radiative transfer package \citep{Narayanan21}, and fitted by the \textsc{prospector} SED model \citep{Johnson19, PROSPECTOR}. They found that, at $z\sim7$, outshining by young stellar populations significantly affects recovered stellar masses. Using synthetic SEDs from simulations provides a `ground truth' to which the SED-recovered physical parameters can be directly compared. In this way, the accuracy of the SED code is determined not in relation to other codes but to the `ground truth' itself. That is the approach taken in this paper.

Here, we investigate whether the \magphys\ model can accurately recover the physical parameters of simulated galaxies from their synthetic SEDs by using a large sample of galaxies from the \eagle\ simulations, for which UV-to-IR SEDs were computed by \cite{Camps16, Camps18} using the \skirt\ radiative transfer code. The \eagle\ simulations provide us with a much larger sample of galaxies (compared to the study of \citealt{Hayward15}) that cover a wider parameter space, which allows us to test \magphys\ more widely and investigate more thoroughly any potential systematic uncertainties and, most importantly, what causes them. For each galaxy, we have an `observed' SED from \skirt\ that we can then fit with \magphys, and we know the `true' physical parameters of these galaxies, that we can compare with the ones recovered by \magphys.
Thus, we enhance the statistical power of previous studies by using a sample of $30,928$ simulated galaxies in three redshift snapshots, $z=0.10$, $z=2.01$, and $z=5.04$. We begin by testing \magphys\ under idealised conditions, fitting the full ultraviolet-to-submillimetre SEDs computed by \skirt\ in fifty photometric bands. We then assess the performance of \magphys\ in more realistic observational scenarios by constructing a mock galaxy sample with photometric coverage and uncertainties matched to those anticipated for the upcoming Wide-Area VISTA Extragalactic Survey (WAVES; \citealt{WAVES}), thereby evaluating the reliability of parameter recovery in a survey-like context.

We note that any such tests are necessarily limited by the fidelity of the simulations used as `ground truth', in particular by the assumptions and resolution limitations inherent to radiative transfer modelling. Our aim is not to claim that \eagle+\skirt\ provides a perfect representation of real galaxies, but rather to use it as a controlled testbed in which the intrinsic physical parameters are known. This enables us to isolate the impact of SED fitting assumptions on parameter recovery in a way that is not possible using observational data alone. In this work, we restrict our analysis to star-forming galaxies and do not consider systems dominated by active galactic nuclei (AGN).

This paper is organised as follows. In Sect.~\ref{sect: Methods}, we describe the \eagle\ simulations, the \skirt\ radiative transfer code, the simulated galaxy samples and the \magphys\ model. In Sect.~\ref{sect: Results}, we present the comparisons between the true physical parameters of the simulated galaxies and the parameters recovered by \magphys. In Sect.~\ref{sect: Discussion} we discuss these results as well as the systematic uncertainties we discover, their potential sources, and ways to reduce them. We summarise and conclude our paper in Sect.~\ref{sect: Conclusion}. Additional results are presented in Appendix~\ref{App: All Old Ages APP}. Throughout this paper, we use AB magnitudes \citep{AB_Mags}, and assume the standard cosmological parameters from Planck \citep{PlanckI, PlanckXVI}, with the key parameters of $\Omega_{m} = 0.307$, $\Omega_{\Lambda} = 0.693$, $\Omega_{b} = 0.04825$, $h = 0.6777$, and $\sigma_{8} = 0.8288$.

\section{Methods} \label{sect: Methods}

\subsection{The Mock Galaxy Sample}

\subsubsection{The \eagle+\skirt\ Simulated Galaxies}
We use a sample of mock galaxies from the Evolution and Assembly of GaLaxies and their Environments (\eagle) simulations that were post-processed by \cite{Camps16} using the Stellar Kinematics Including Radiative Transfer (\skirt) radiative transfer code, and released as a database of UV to submillimetre broadband fluxes by \cite{Camps18}.

\textsc{eagle} \citep{Crain15,Schaye15} is a suite of hydrodynamical simulations that follow the formation of large-scale structure and galaxies in a $\Lambda$CDM Universe. The simulations were run using a modified version of the \textsc{gadget-3} smooth particle hydrodynamical (SPH) code \citep{Gadget05}. This modified hydrodynamics scheme, known as ANARCHY \citep{Schaye15,Schaller15}, affects the implementation of SPH, time-stepping, and subgrid models of the simulations that govern cooling, star formation, stellar evolution, feedback from star formation, gas accretion on to and mergers of black holes, and active galactic nucleus (AGN) feedback. Within these simulations, galaxies are defined as gravitationally-bound sub-halos identified through use of the friends-of-friends (FoF) algorithm \citep{Davis85} and the SUBFIND algorithm \citep{Gadget01,Dolag09}. Galaxy properties are then calculated from the relevant particles in the corresponding sub-halo and that are within a 3D aperture with a radius of 30 proper kpc (pkpc) from the galaxy's stellar centre of mass.
The subgrid physics used in the \eagle\ simulations is based on the physics implemented in the OverWhelmingly Large (OWL) Simulations \citep{Schaye10}, and is described in detail in \cite{Schaye15}. The \eagle\ simulations provide a realistic set of simulated galaxies, as they have been shown to reproduce the stellar mass functions and optical colours of local galaxies \citep{Furlong15,Trayford15}. Following the description of the \eagle\ simulations in \citet{Schaye15} and \citet{Crain15}, \citet{McAlpine16} presents a database of these simulations for public use.

\textsc{skirt} is a 3D Monte Carlo radiative transfer code that simulates the transfer of radiation through dusty media  \citep{SKIRT03,SKIRT11,SKIRT15}. It follows the lives of photons as they travel through the interstellar medium (ISM), where they are absorbed or scattered by dust, and computes the heating of and re-emission by dust grains self-consistently. It can be used to consistently model the total stellar and dust emission of galaxies (e.g., \citealt{Baes10, deLooze2012}). Later updates to \skirt\ allow the code to self-consistently calculate the dust emission of galaxies, assuming the dust content of said galaxies are in local thermal equilibrium (LTE; \citealt{Baes05a, Baes05b}). Early versions of the \skirt\ code used the \citet{Zubko04} dust model only to describe the dust optical properties of galaxies \citep{Baes10}, however later updates have added multiple different dust models that can be chosen to use, including the \citet{Draine07} dust model which is built into the DustME library \citep{Compiegne11}. By linking the \skirt\ code to DustME, a numerical tool designed to calculate the emission of dust assuming LTE, the \skirt\ code also calculates the dust temperature distribution of a galaxy and the associated FIR and submm emission, including that from transiently heated dust grains and PAHs molecules \citep{SKIRT11}. This is possible even for galaxies with various custom mixtures of dust grains, dust distributions, and geometry of stellar components, which can all be selected and combined in as the user desires \citep{SKIRT15,Baes15}.

\cite{Camps16} used \skirt\ to post-process a sample of \eagle\ galaxies with properties that match the selection criteria of local dusty galaxies from the {\it Herschel} Reference Survey \citep{Boselli10,Cortese12}. For the \skirt\ processing, they selected the ZubkoDustMix dust mixture \citep{Zubko04} and the OctTreeDustGrid spatial distribution of the dusty medium \citep{Saftly13}. They overcame the resolution limitations of the simulations by employing sub-grid models for the star-forming regions and for the diffuse dust distribution in the \eagle\ galaxies which are thoroughly detailed in \citet{Camps16}. \skirt\ produces the UV to submm emission for each galaxy in different photometric bands to obtain fluxes that can be directly compared with observations of galaxies using various telescopes and instruments. \cite{Camps16} then used these mock observations to show that \eagle\ can reproduce the infrared and submm properties of local galaxies.

\cite{Camps18} later extended this work to a larger sample of nearly half a million simulated \eagle\ galaxies with $\log(\mstar/\msun)>8.5$ and sufficiently resolved dust (with the number of numerical particles representing the dust content in a galaxy $N_\mathrm{dust}>250$) in 23 redshift snapshots up to redshift $z=6$, across six \eagle\ runs. They found 316,389 galaxies that produced sufficiently resolved dust distributions to calculate meaningful dust-attenuated stellar emission and dust emission fluxes. The galaxies retain their original viewing angles from \eagle, which are labelled as `random' viewing angles, as \citet{Camps18} also considers edge-on and face-on viewing angles at UV and optical wavelengths. They then extended the \eagle\ database by publishing the rest-frame magnitudes and observer-frame fluxes of these 316,389 galaxies in 50 standard photometric filters from UV to mm wavelengths ($\sim$$0.15$ to $\sim$$1250$ $\mu$m). These bands are: GALEX FUV/NUV \citep{GALEX_bands}; SDSS {\it ugriz} \citep{SDSS_bands}; 2MASS {\it JHKs} \citep{2MASS_bands}; UKIDSS {\it ZYJHK} \citep{UKIDSS_bands}; Johnson UBVRIJM \citep{Johnson_bands}; WISE W1/W2/W3/W4 \citep{WISE_bands}; IRAS 12/25/60/100 $\mu$m \citep{IRAS_bands}; {\it Spitzer}/IRAC I1/I2/I3/I4 \citep{IRAC_bands}; {\it Spitzer}/MIPS 24/70/160 $\mu$m \citep{MIPS_bands}; {\it Herschel}/PACS 70/100/160 $\mu$m \citep{PACS_bands}; {\it Herschel}/SPIRE 250/350/500 $\mu$m \citep{SPIRE_bands}; SCUBA2 450/850 $\mu$m \citep{SCUBA2_bands}; and ALMA Bands 10/9/8/7/6 \citep{ALMA_bands}. The names and pivot wavelengths of these photometric filters are displayed in Table~\ref{tab: 50 filters}, and ordered from shortest to longest wavelength. These magnitudes and fluxes are given in the `DustyMagnitudes' and `DustyFluxes' tables of the database, respectively. The `DustFit' table was also added to the database to provide the computed dust temperature and dust mass of each galaxy.

\begin{table}
    \centering
    \begin{tabular}{| lr |}
        \hline
        Filter name & $\lambda$\textsubscript{pivot} ($\mu$m) \\
        \hline
        GALEX\_FUV & 0.1535 \\
        GALEX\_NUV & 0.2301 \\
        Johnson\_U & 0.3525 \\
        SDSS\_u & 0.3557 \\
        Johnson\_B & 0.4417 \\
        SDSS\_g & 0.4702 \\
        Johnson\_V & 0.5525 \\
        SDSS\_r & 0.6176 \\
        Johnson\_R & 0.6899 \\
        SDSS\_i & 0.7490 \\
        Johnson\_I & 0.8739 \\
        UKIDSS\_Z & 0.8826 \\
        SDSS\_z & 0.8947 \\
        UKIDSS\_Y & 1.031 \\
        TwoMASS\_J & 1.239 \\
        Johnson\_J & 1.243 \\
        UKIDSS\_J & 1.250 \\
        UKIDSS\_H & 1.649 \\
        TwoMASS\_H & 1.649 \\
        TwoMASS\_Ks & 2.164 \\
        UKIDSS\_K & 2.206 \\
        \hline
    \end{tabular}
    \quad
    \begin{tabular}{| lr |}
        \hline
        Filter name & $\lambda$\textsubscript{pivot} ($\mu$m) \\
        \hline
        WISE\_W1 & 3.390 \\
        IRAC\_I1 & 3.551 \\
        IRAC\_I2 & 4.496 \\
        WISE\_W2 & 4.641 \\
        Johnson\_M & 5.012 \\
        IRAC\_I3 & 5.724 \\
        IRAC\_I4 & 7.884 \\
        IRAS\_12 & 11.41 \\
        WISE\_W3 & 12.57 \\
        WISE\_W4 & 22.31 \\
        IRAS\_25 & 23.61 \\
        MIPS\_24 & 23.76 \\
        IRAS\_60 & 60.41 \\
        PACS\_70 & 70.77 \\
        MIPS\_70 & 71.99 \\
        PACS\_100 & 100.8 \\
        IRAS\_100 & 101.1 \\
        MIPS\_160 & 156.4 \\
        PACS\_160 & 161.9 \\
        SPIRE\_250 & 252.5 \\
        ALMA\_10 & 349.9 \\
        SPIRE\_350 & 354.3 \\
        SCUBA2\_450 & 449.3 \\
        ALMA\_9 & 456.2 \\
        SPIRE\_500 & 515.4 \\
        ALMA\_8 & 689.6 \\
        SCUBA2\_850 & 853.8 \\
        ALMA\_7 & 937.9 \\
        ALMA\_6 & 1244 \\
        \hline
    \end{tabular}
    \caption[The 50 photometric bands included in the \eagle\ database]{The names and pivot wavelengths of the 50 parametric bands for which mock broadband observed fluxes and absolute AB magnitudes are available in the \eagle\ database. The left table list the UV and optical filters, for which face-on, edge-on and random orientation measurements are available. The right table lists the IR and submm filters which only have one orientation available, as emission in these bands is essentially isotropic.}
    \label{tab: 50 filters}
\end{table}

\subsubsection{The idealised subsample of mock galaxies} \label{mock_sample}

To build our idealised mock galaxy sample, we use star-forming galaxies from three redshift snapshots provided by the \eagle\ database: snapshots 27 ($z=0.10$), 15 ($z=2.01$) and 8 ($z=5.04$); we chose these snapshots as representative of the low-, intermediate-, and high-redshift Universe, respectively. We selected galaxies from five of the six available \eagle\ models in the database: Ref-L0025N0752, Recal-L0025N0752, Ref-L0025N0376, Ref-L0050N0752, and Ref-L0100N1504. These models differ in: subgrid physics assumptions, indicated by the `Ref' or `Recal' prefix; box-length of the simulation, with each box-length given in comoving Mpc by the four digits following `L'; and total number of both the baryonic and dark matter particles, the cubed root of which is given by the four digits following `N'. We exclude galaxies from the sixth \eagle\ model, AGNdT9-L0050N0752, as we are only interested in testing the \magphys\ model without AGN templates for now. This allows us to create three samples of mock galaxies at three redshifts. The distribution of these galaxies by model and redshift is shown in Table~\ref{tab: Large Sample}.

\begin{table}
    \centering
    \begin{tabular}{| l | ccc |}
        \hline
        \eagle\ Model & Galaxies at & Galaxies at & Galaxies at \\
         & $z=0.10$ & $z=2.01$ & $z=5.04$ \\
        \hline
        Ref-L0025N0752 & 527 & 298 & 28 \\
        Recal-L0025N0752 & 384 & 233 & 26 \\
        Ref-L0025N0376 & 155 & 227 & 18 \\
        Ref-L0050N0752 & 1150 & 2003 & 223 \\
        Ref-L0100N1504 & 8072 & 15445 & 2139 \\
        \hline
        Combined Total & 10288 & 18206 & 2434 \\
        \hline
    \end{tabular}
    \caption[The \eagle\ galaxies included in our samples]{The \eagle\ galaxies that are included in our three samples used when testing the \magphys\ model, displayed by \eagle\ model and redshift. In total, we investigated $30,928$ galaxies.}
    \label{tab: Large Sample}
\end{table}

We obtained the SEDs and parameters of these galaxies from the \eagle\ database. Specifically from \citet{Camps18} we obtained:

\begin{itemize}
    \item the observer-frame, dust-affected fluxes at 50 wavelength filters ranging from $\sim$$0.15$ to $\sim$$1250$ $\mu$m from the `DustyFluxes' table;
    \item and the representative dust temperatures and dust masses, provided as `Temp\_Dust' and `Mass\_Dust' on the `DustFit' table.
\end{itemize}

\noindent From \citep{McAlpine16} we obtained:
\begin{itemize}
    \item the average, birth mass-weighted stellar population ages, provided as `InitialMassWeightedStellarAge' on the `SubHalo' table;
    \item the total current masses (baryonic and dark matter), the gas masses, and stellar masses, provided as `Mass', `MassType\_Gas' and `MassType\_Star' on the `SubHalo' table;
    \item the total star formation rates, provided as `StarFormationRate' on the `SubHalo' table;
    \item the stellar metallicities, provided as `Stars\_Metallicity' on the `SubHalo' table;
    \item and the stellar half-mass radii within an aperture of 30 pkpc, provided as `R\_halfmass30' on the `Sizes' table.
\end{itemize}

\begin{figure*}
    \centering
    \includegraphics[width=\linewidth]{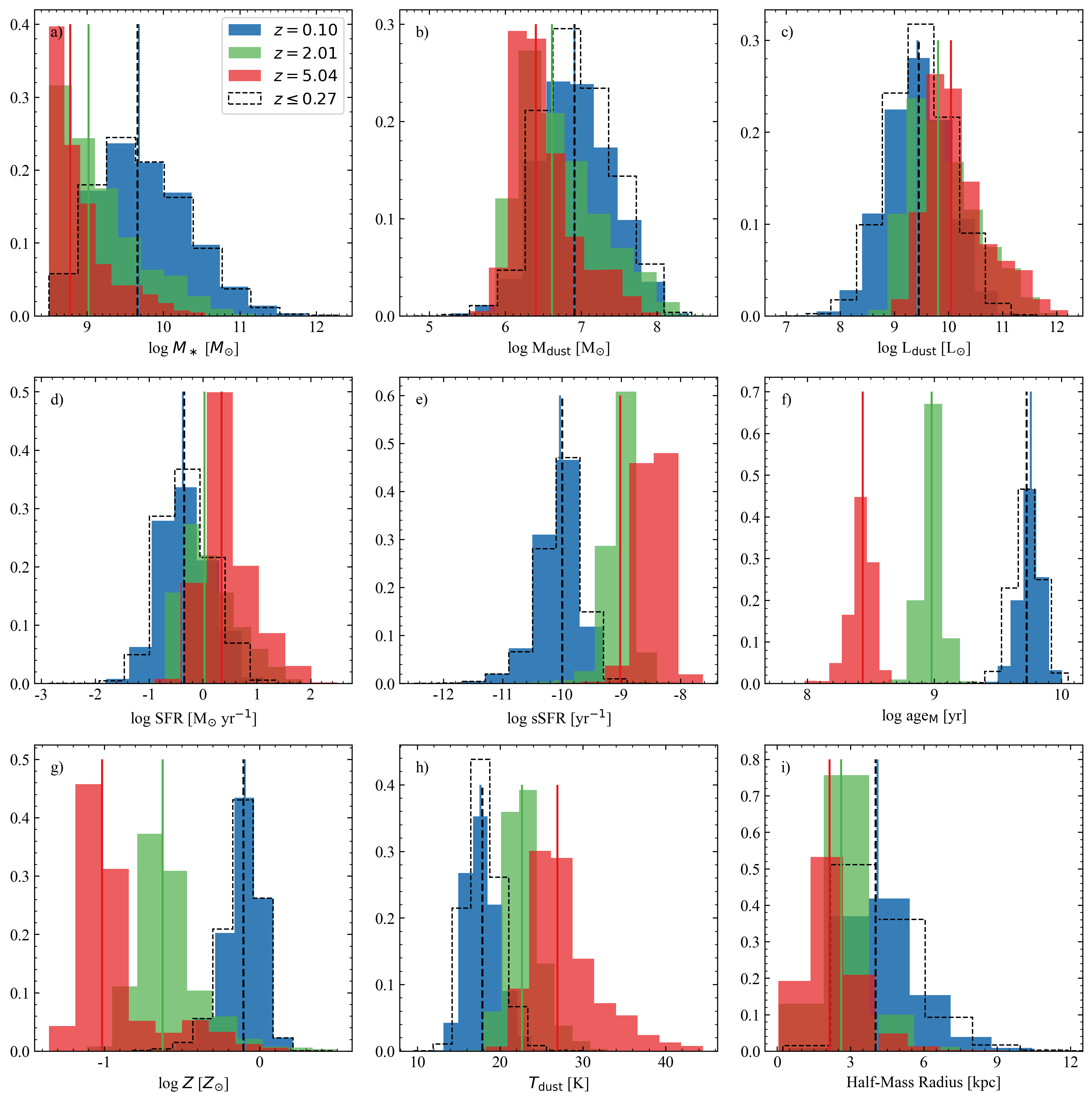}
    \caption{Normalised distributions of nine parameters of the 10,288 galaxies at redshift $z=0.10$ (shown in blue), 18,206 galaxies at redshift $z=2.01$ (shown in green), and 2,434 galaxies at redshift $z=5.04$ (shown in red) of our idealised mock galaxy sample, and 35,051 galaxies at redshift $z \leq 0.27$ (shown in black) of our WAVES-like sample. The nine parameters displayed are: (a) stellar mass, (b) dust mass, (c) dust luminosity, (d) SFR, (e) sSFR, (f) average stellar age, (g) stellar metallicity, (h) dust temperature, and (i) half-mass radius. The vertical lines show the median of each distribution.}
    \label{fig: EAGLE hists}
\end{figure*}

\begin{figure}
    \centering
    \includegraphics[width=\columnwidth]{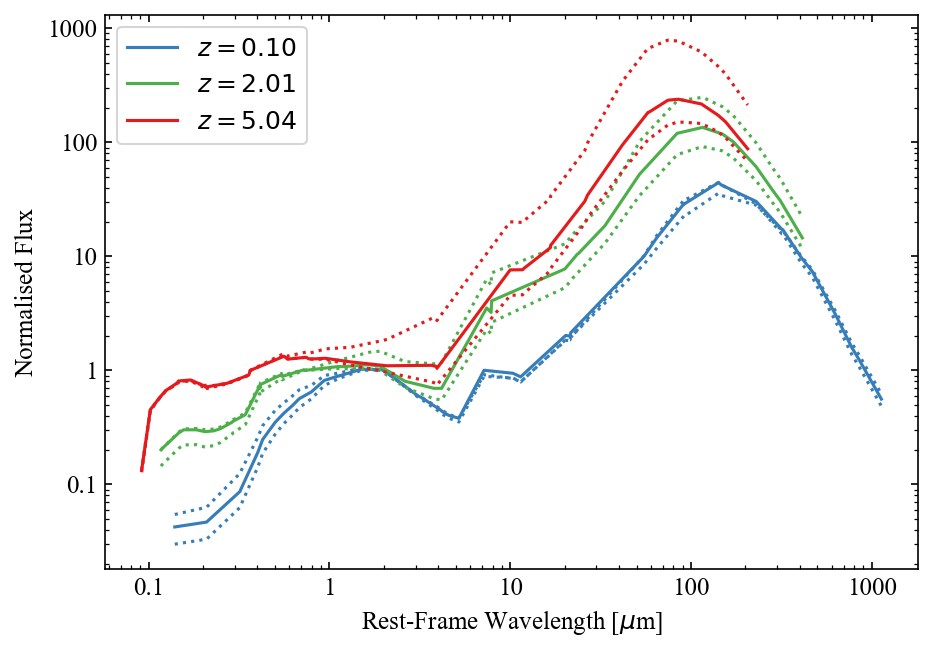}
    \caption{Median synthetic SEDs of the three redshift samples (normalised in the $K$-band), showing the 16\textsuperscript{th} and 84\textsuperscript{th} percentile values as dotted lines, and the 50\textsuperscript{th} percentile values as solid lines. The redshift $z=0.10$ galaxies are shown in blue, the redshift $z=2.01$ galaxies are shown in green, and the redshift $z=5.04$ galaxies are shown in red.}
    \label{fig: EAGLE SEDs}
\end{figure}

Fig.~\ref{fig: EAGLE hists} displays the normalised distributions of nine key parameters of the galaxies in our mock sample: stellar mass, dust mass, dust luminosity, SFR, sSFR, average stellar age, stellar metallicity, dust temperature and half-mass radius. Our local ($z=0.1$) mock galaxy sample includes higher stellar mass galaxies with higher metallicities, older ages, and lower specific star formation rates, as expected from a more evolved population of galaxies in the local Universe. The high-redshift ($z=5.04$) sample extends to lower masses and metallicities, younger ages, and higher specific star formation rates. We note the very narrow range of average stellar ages in each of the mock galaxy subsamples (panel f): the subsamples do not overlap in ages at all. While the stellar ages in galaxies at higher redshifts are limited by the age of the Universe at those redshifts, younger average ages are possible at low redshifts if there is significant recent star formation; we will return to this issue in Section~\ref{sect: Discussion}.

We also display the $K$-band normalised median SEDs for each of the three subsamples in Fig.~\ref{fig: EAGLE SEDs}, which shows the average shape of the synthetic SEDs. We note the relative lack of variability of the synthetic SEDs in the $z=0.1$ sample compared to the higher redshift subsamples. On average, the local SEDs display redder UV/optical colours, as a result of their relatively older ages combined with dust attenuation. Their infrared SEDs peak at longer wavelengths as a result of the colder dust temperatures, as shown in Fig.~\ref{fig: EAGLE hists} (h). It is interesting to note that the mock galaxies seem to display much larger variation of their optical-to-IR SEDs in the highest-redshift subsample.

\subsubsection{The WAVES-like subsample of mock galaxies} \label{waves_sample}

To evaluate the performance of \magphys\ under realistic observational conditions, we construct an additional subsample of mock galaxies designed to mimic the properties of the upcoming Wide-Area VISTA Extragalactic Survey (WAVES; \citealt{WAVES}). WAVES will obtain spectroscopy for approximately $1.6$ million galaxies using the 4-metre Multi-Object Spectroscopic Telescope (4MOST; \citealt{4MOST}), divided into two components: WAVES Wide and WAVES Deep. In this work, we focus on the WAVES Wide survey, which aims to obtain spectra for approximately $0.9$ million galaxies over an area of $\sim1200$ square degrees.

For the majority of WAVES Wide galaxies, broad-band \textit{ugriZYJHK} photometry will be available from the KiDS \citep{Wright24} and VIKING \citep{Edge13} imaging surveys. The survey primarily targets galaxies at redshifts $z \lesssim 0.2$, with a limiting $Z$-band magnitude of $Z \leq 21.1$.

To reproduce the redshift range probed by WAVES Wide, we extend our idealised low-redshift sample by including additional \eagle\ snapshots at $z = 0.00$, $0.18$, and $0.27$, drawn from the same five \eagle\ simulation runs used previously. Including galaxies up to $z = 0.27$ allows us to capture the effective redshift range expected for WAVES Wide once photometric redshift uncertainties are taken into account. This increases the size of the low-redshift mock sample from $10,288$ galaxies at $z = 0.10$ to $42,588$ galaxies at $z \leq 0.27$.

We then apply a $Z$-band magnitude cut of $Z \leq 21.1$, matching the WAVES Wide selection criteria, which yields a final WAVES-like sample of $35,051$ galaxies. The physical properties of these mock galaxies are plotted in Fig.~\ref{fig: EAGLE hists}. For this sample, we fit only the nine \textit{ugriZYJHK} photometric bands available to WAVES, rather than the full set of 50 bands provided in the \eagle\ database.

To assign realistic photometric uncertainties, we adopt the limiting magnitudes reported in Table~1 of \citet{Wright24} and convert them into flux uncertainties assuming Gaussian errors. This WAVES-like subsample therefore enables us to assess the accuracy and robustness of \magphys\ parameter recovery in a survey-like observational context.


\subsection{SED fitting with \magphys}
The \magphys\ SED modelling code \citep{magphys, daCunha15} can interpret the UV, optical, and near-infrared emission from stellar populations in galaxies consistently with their dust emission in the infrared regime. \magphys\ uses the stellar population synthesis models of \citet{Bruzual03} to compute light emitted by stellar populations of various ages and metallicities in galaxies between $91${\AA} and $160\mu$m. The \cite{Chabrier03} IMF is assumed. The stellar emission from galaxies is computed by combining these stellar populations with a range of star formation histories that rise linearly at early ages and then decline exponentially, with varying timescales and superimposed random bursts of star formation to account for stochasticity in the star formation histories. This starlight is then attenuated by dust using the two-component model of \citet{Charlot00}, which accounts for newly-formed stars ($<10$\,Myr) inside their birth clouds being more attenuated than older stars in the diffuse ISM. The effective dust attenuation curves in the birth cloud and diffuse ISM components have different slopes: these are steeper in the birth cloud component ($\tau_\lambda \propto \lambda^{-1.3}$) than in the diffuse ISM ($\tau_\lambda \propto \lambda^{-0.7}$). This produces an age-dependent effective dust attenuation curve (see more details in \citealt{Charlot00, magphys, Battisti20}).

By combining the emission from the stellar populations with this dust attenuation model, \magphys\ computes the total amount of energy absorbed by dust in both stellar birth clouds and the diffuse ISM and, through an energy balance argument, assumes that emission must be re-radiated by dust grains in the mid- to far-infrared. This emission is modelled using an empirical template for the polycyclic aromatic hydrocarbon (PAH) mid-infrared features, a hot dust component for the mid-infrared continuum, and two modified black bodies for the far-infrared continuum, which have dust equilibrium temperatures that are kept as free parameters. The relative contributions of these different dust emission components to the total dust emission are also free parameters.

To fit the observed SEDs of galaxies, \magphys\ generates a random library of stellar population models, for various star formation histories, metallicities and dust content, and a random library of dust emission models, for various dust temperatures and contributions by different dust components. These libraries are designed to encompass all plausible parameter combinations of the observed SED. The final stellar population library and the final infrared spectra library both consist of 50,000 different models each. The models in the two libraries are associated with each other through the parameter $f_{\mu}$, which is the fraction of the total infrared luminosity contributed by dust in the ISM. For each association with a sufficiently similar $f_{\mu}$ value ($\Delta f_\mu \leq 0.15$), the combined stellar and infrared spectra is included in the final model SED library.

The final library of combined UV, optical, and IR SEDs consists of $\sim$$661$ million models. This large number of models is required to properly sample the multi-dimension observational space. \magphys\ compares the observed flux in each photometric band with the flux predicted by each model in the library and computes the $\chi^2$ goodness-of-fit and probability of each model. Then, for each parameter, it marginalises the probability over all other parameters to produce a posterior likelihood distribution. We adopt the \magphys\ estimated parameter value to be the median of that likelihood distribution, and the confidence range to be its 16th-84th percentile range.

We formatted the flux measurements for each galaxy described in Sect.~\ref{mock_sample} when observed at random viewing orientations as \magphys\ inputs. In total, we fit the UV-to-submm SEDs of 30,928 mock galaxies at redshifts $z=0.10$, $z=2.01$, and $z=5.04$ from the \eagle\ Ref-L0025N0752, Recal-L0025N0752, Ref-L0025N0376, Ref-L0050N0752, and Ref-L0100N1504 simulations. We did not add noise to the photometric data, and instead applied a uniform uncertainty of 20\% to each flux measurement, in order to investigate potential systematic uncertainties present within \magphys\ rather than those caused by noisy data.


For the redshift $z=0.10$ subsample, we use the \magphys\ priors described in \cite{magphys}, whereas for the redshift $z=2.01$ and $z=5.04$ subsamples, we use the \magphys-highz version \citep{daCunha15}, which includes priors that are more appropriate for high-redshift galaxies (specifically, a more diverse range of SFHs, and an extension to higher dust attenuations and warmer dust temperatures). For each mock galaxy, \magphys\ fits the input SED and returns a best-fit model along with the likelihood distributions of key parameters. Here we focus on the following parameters:

\begin{itemize}
\item stellar mass, \mstar;
\item star formation rate (averaged over the last 100 Myr), SFR;
\item specific star formation rate, ${\rm sSFR}={\rm SFR}/\mstar$;
\item total dust luminosity, \ldust;
\item total dust mass, \mdust;
\item mass-weighted age, \agem.
\end{itemize}

We compare these parameters directly with the `true' values provided by \eagle+\skirt\ in Sect.~\ref{sect: Results}. In some cases, the \eagle+\skirt\ database does not provide the parameter of interest, or the \magphys\ and \eagle+\skirt\ parameters are defined in slightly different ways. We discuss these cases in detail when presenting their results.

\subsection{Comparison metrics}
In our analysis, we quantify the accuracy of the recovered \magphys\ parameter values by measuring the median difference or offset between them and the true \eagle\ values, $\Delta$, given as:
\begin{equation}
    \Delta = \text{median}(x_{\text{\magphys}} - x_{\text{\eagle}})\,,
\end{equation}
where $x_{\text{\magphys}}$ and $x_{\text{\eagle}}$ correspond to the recovered logarithmic \magphys\ values and true logarithmic \eagle\ values, respectively. We additionally find for each parameter the standard deviations, lower quartiles (Q1), upper quartiles (Q3), median error bars, and normalised median absolute deviation (NMAD) of each parameters, with NMAD defined as:

\begin{equation}
    \text{NMAD} = 1.4826 \times \text{median}(|(x_{\text{\magphys}} - x_{\text{\eagle}}) - \Delta|)\,,
\end{equation}
where NMAD provides a robust estimate of the scatter around the median trend, approximately equivalent to the standard deviation for a Gaussian distribution but less sensitive to outliers.

Finally, we find the outlier fraction of each parameter, where an outlier value is either lower than $Q1 - 1.5 \times IQR$ or higher than $Q3 + 1.5 \times IQR$, and $IQR$ is the interquartile range.

\section{Results} \label{sect: Results}

\begin{table*}
    \centering
    \resizebox{\textwidth}{!}{
    \begin{tabular}{| l | l | rrrrrrr |}
        \hline
        Parameter & \eagle\ Database Name & Median Offset & Standard Deviation & NMAD & Offset Q1 & Offset Q3 & Outlier Fraction & Median Error Bar \\
        \hline
        \multicolumn{9}{|c|}{Redshift $z=0.10$ Subsample} \\
        \hline
        $\log(\mstar/\msun)$ & MassType\_Star & -0.140 & 0.078 & 0.066 & -0.190 & -0.099 & 0.031 & 0.210 \\
        $\log(\rm{SFR}/\msun\,\rm{yr}^{-1})$ & StarFormationRate & -0.064 & 0.110 & 0.094 & -0.132 & -0.005 & 0.025 & 0.115 \\
        $\log(\rm{sSFR}/\rm{yr}^{-1})$ & - & 0.072 & 0.138 & 0.134 & -0.016 & 0.165 & 0.012 & 0.270 \\
        $\log(\agem/\rm{yr})$ & InitialMassWeightedStellarAge & -0.256 & 0.143 & 0.121 & -0.356 & -0.183 & 0.029 & 0.345 \\
        $\log(\ldust/L_\odot)$ & - & 0.014 & 0.038 & 0.035 & -0.010 & 0.037 & 0.024 & 0.079 \\
        $\log(\mdust/\msun)$ & Mass\_Dust & 0.071 & 0.058 & 0.049 & 0.036 & 0.103 & 0.020 & 0.090 \\
        \hline
        \multicolumn{9}{|c|}{Redshift $z=2.01$ Subsample} \\
        \hline
        $\log(\mstar/\msun)$ & MassType\_Star & -0.050 & 0.142 & 0.066 & -0.101 & -0.010 & 0.106 & 0.075 \\
        $\log(\rm{SFR}/\msun\,\rm{yr}^{-1})$ & StarFormationRate & -0.027 & 0.098 & 0.058 & -0.065 & 0.013 & 0.047 & 0.045 \\
        $\log(\rm{sSFR}/\rm{yr}^{-1})$ & - & 0.022 & 0.165 & 0.109 & -0.045 & 0.105 & 0.067 & 0.110 \\
        $\log(\agem/\rm{yr})$ & InitialMassWeightedStellarAge & -0.096 & 0.172 & 0.135 & -0.193 & -0.010 & 0.041 & 0.120 \\
        $\log(\ldust/L_\odot)$ & - & -0.066 & 0.033 & 0.016 & -0.078 & -0.055 & 0.062 & 0.025 \\
        $\log(\mdust/\msun)$ & Mass\_Dust & 0.002 & 0.071 & 0.059 & -0.041 & 0.040 & 0.041 & 0.160 \\
        \hline
        \multicolumn{9}{|c|}{Redshift $z=5.04$ Subsample} \\
        \hline
        $\log(\mstar/\msun)$ & MassType\_Star & -0.070 & 0.203 & 0.114 & -0.163 & -0.002 & 0.109 & 0.035 \\
        $\log(\rm{SFR}/\msun\,\rm{yr}^{-1})$ & StarFormationRate & -0.019 & 0.145 & 0.073 & -0.075 & 0.026 & 0.081 & 0.020 \\
        $\log(\rm{sSFR}/\rm{yr}^{-1})$ & - & 0.059 & 0.210 & 0.148 & -0.034 & 0.171 & 0.045 & 0.030 \\
        $\log(\agem/\rm{yr})$ & InitialMassWeightedStellarAge & -0.112 & 0.223 & 0.148 & -0.259 & -0.031 & 0.073 & 0.020 \\
        $\log(\ldust/L_\odot)$ & - & -0.056 & 0.041 & 0.026 & -0.074 & -0.039 & 0.044 & 0.020 \\
        $\log(\mdust/\msun)$ & Mass\_Dust & -0.002 & 0.078 & 0.047 & -0.031 & 0.034 & 0.085 & 0.270 \\
        \hline
    \end{tabular}}
    \caption[Summary of the six galaxy properties investigated]{The six parameters we investigated to determine the accuracy of the \magphys\ model. The columns show parameter names and units, parameter names used in the \eagle\ database (if they are provided), median offsets, standard deviations, NMAD, Q1 offsets, Q3 offsets, outlier fractions, and median error bars of the recovered \magphys\ values when compared to the true \eagle\ values. Our sample has been separated by redshift subsamples for ease of comparison.}
    \label{tab: parameter summary}
\end{table*}

\subsection{Quality of the fits}

\begin{figure}
    \centering
    \includegraphics[width=\columnwidth]{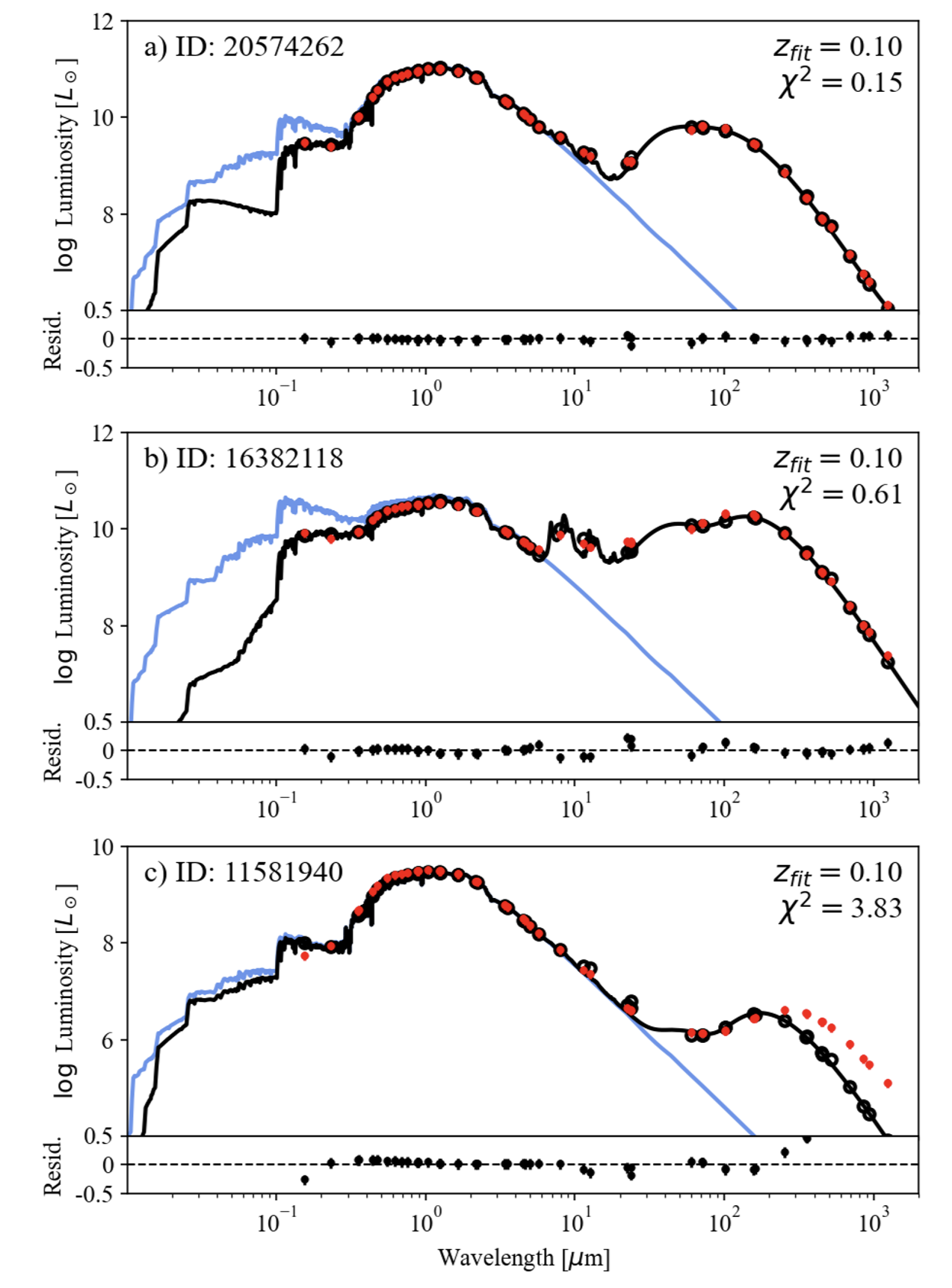}
    \caption[Three SEDs fit by \magphys]{The SEDs of three of the \eagle\ galaxies when fit by \magphys. The observed fluxes at each wavelength are shown as red dots, the unattenuated stellar emission are shown in blue, and the SEDs are shown in black, with the predicted fluxes at each wavelength shown as open, black circles. The galaxy ID is shown in the top left corner of each panel. The redshift of the galaxy and the $\chi^{2}$ value of each fit is shown in the top right corner of each panel. The residuals for each fit are shown below the SEDs, and share the same definition as the residuals shown in Fig~\ref{fig: Ave SEDs}. The wavelengths shown are in the observer frame of reference.}
    \label{fig: SED fitting}
\end{figure}

We first check how well \magphys\ fits the synthetic SEDs provided by \eagle. Fig.~\ref{fig: SED fitting} presents the SEDs of three of the \eagle\ galaxies at redshift $z=0.10$. The \eagle\ galaxy with ID 20574262, shown in panel (a), was the best-fit galaxy of our redshift $z=0.10$ subsample, with the lowest $\chi^2$. Most of the observed fluxes for this galaxy lie exactly on the curve predicted by \magphys, which indicates the high quality of this fit. Panel (b) displays the fit for \eagle\ galaxy 16382118, which has a fit $\chi^2$ value close to the median of the whole low-$z$ subsample. This is still a very good fit, as shown by the small residuals across wavelength. Panel (c) displays the fit for \eagle\ galaxy 11581940, one of the worst-fit galaxies. Even with a relatively large $\chi^{2}$, most of the flux observations of this galaxy are matched very closely to the model SED. The fluxes that are not matched well are primarily in the FIR, which could be attributed to the differences in the dust temperatures of the models: Fig.~\ref{fig: EAGLE hists} shows that some \eagle\ galaxies have dust temperatures below 15~K, which is the coldest dust temperature allowed by the \magphys\ prior.

\begin{figure}
    \centering
    \includegraphics[width=\columnwidth]{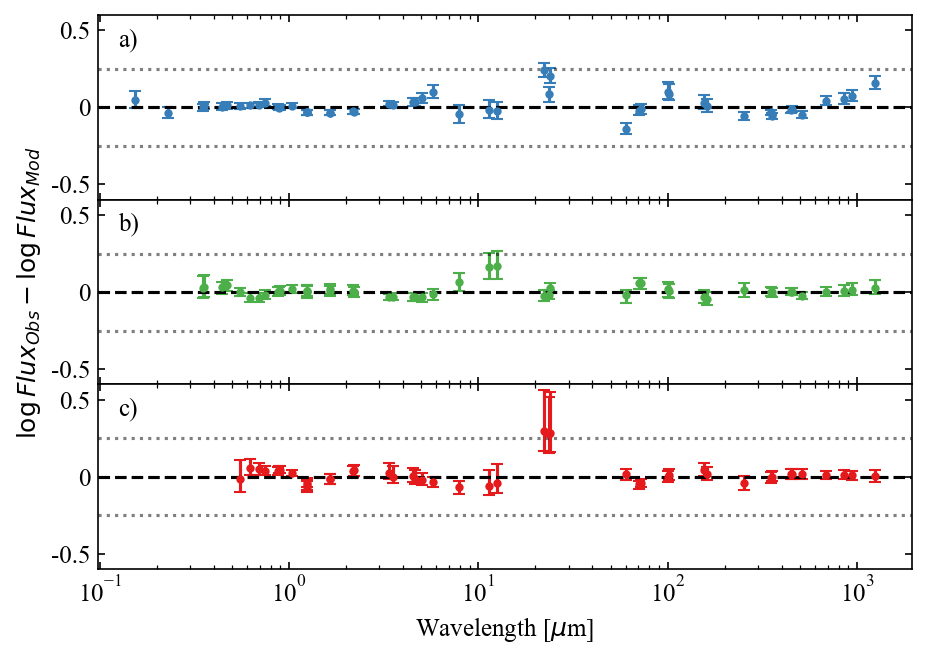}
    \caption[Median SED residuals for three redshift samples]{The median SED residuals for the three redshift samples, with redshift $z=0.10$ galaxies shown in panel a) in blue, redshift $z=2.10$ galaxies shown in panel b) in green, and redshift $z=5.04$ galaxies shown in panel c) in red. The error bars of all three panels show the 16\textsuperscript{th} and 84\textsuperscript{th} percentiles of the residuals, which share the same definition as the residuals shown in Fig~\ref{fig: SED fitting}. The wavelengths shown are in the observer frame of reference.}
    \label{fig: Ave SEDs}
\end{figure}

To investigate how well the SEDs are fitted in general, we plot in Fig.~\ref{fig: Ave SEDs}, the median fit residuals in each band for the three redshift subsamples. The median differences between the observed and model flux values are very low across all wavelengths. The largest residuals are found in the observer-frame MIR regime, which could be due to the relatively simple PAHs emission template used by \magphys\ in that regime. There are also significant residuals in the longest wavelength band at $z=0.10$, which are likely due to the fact that some \eagle\ galaxies extend to colder dust temperatures than allowed by the \magphys\ prior. When assessing the reduced $\chi^{2}$ values of our sample, which compares the flux measurements of the observed spectra to the model spectra at each photometric band, we find that 99.7\% of the redshift $z=0.10$ galaxies, 98.3\% of the redshift $z=2.01$ galaxies, and 73.5\% of the redshift $z=5.04$ galaxies have a $\chi^{2}$ value of $\leq 1$.

\begin{figure*}
    \centering
    \includegraphics[width=\textwidth]{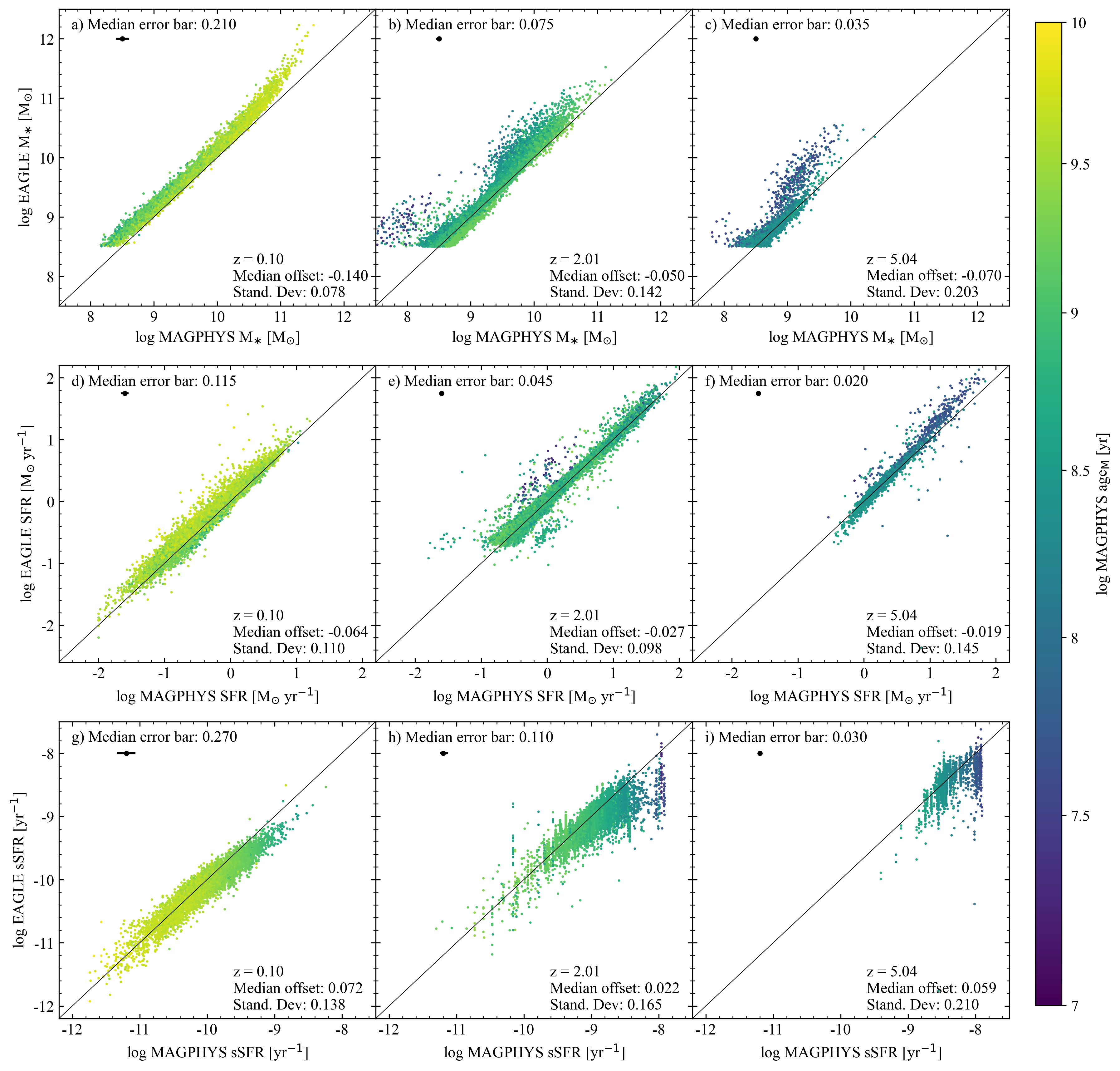}
    \caption[Comparison plots for Stellar Mass, Dust Mass and Dust Luminosity at three redshifts]{Comparisons between the true parameters of the \eagle\ mock galaxies ($y$-axes), and the median-likelihood parameters recovered by \magphys\ from fitting their \skirt-processed SEDs ($x$-axes). The left-hand, middle, and right-hand columns show the comparison for the redshift $z=0.10$, $z=2.01$, and $z=5.04$ subsamples, respectively, for each parameter. The top row shows the comparisons of stellar masses, the middle row shows SFR, and the bottom row shows sSFR. All three parameters are coloured by the recovered \magphys\ mass-weighted average stellar ages. The diagonal black line represents the one-to-one relation. Printed in the bottom right corners of the panels are the median offsets between the trends and the one-to-one relation, and the standard deviations of the trends. Printed in the top left corners of the panels are the median error bars of the recovered values, defined as the median difference between the 16\textsuperscript{th} and 84\textsuperscript{th} percentiles of each recovered value.}
    \label{fig: Params 1}
\end{figure*}

\begin{figure*}
    \centering
    \includegraphics[width=\textwidth]{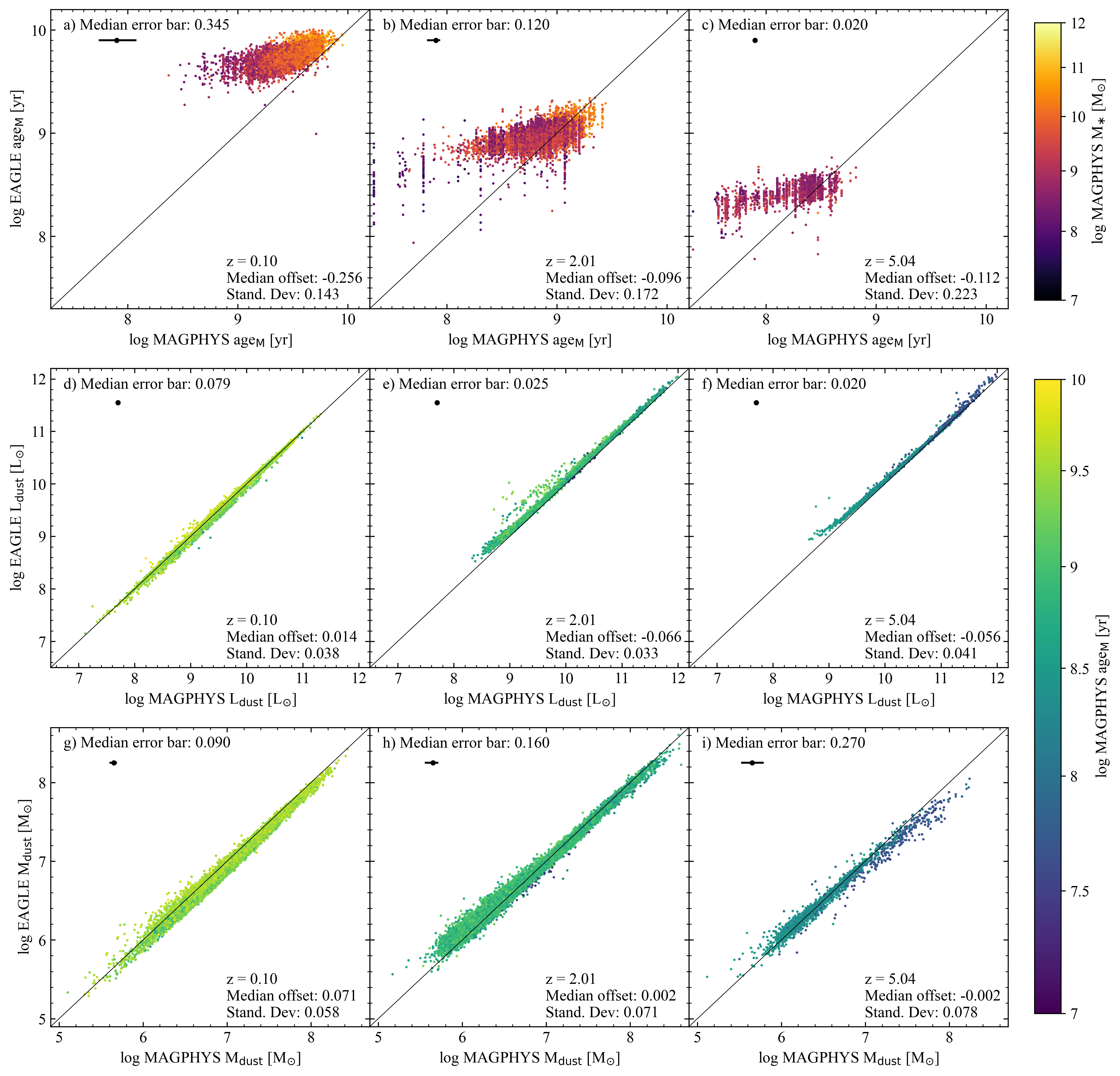}
    \caption[Comparison plots for SFR, sSFR and Average Stellar Age at three redshifts]{Comparisons between the true parameters of the \eagle\ mock galaxies ($y$-axes), and the median-likelihood parameters recovered by \magphys\ from fitting their \skirt-processed SEDs ($x$-axes). The left-hand, middle, and right-hand columns show the comparison for the redshift $z=0.10$, $z=2.01$, and $z=5.04$ subsamples, respectively, for each parameter. The top row shows the comparisons of average stellar ages, the middle row shows dust luminosities, and the bottom row shows dust masses. Average stellar age is coloured by the recovered \magphys\ stellar masses, and dust luminosity and mass are coloured by the recovered \magphys\ mass-weighted average stellar ages. The diagonal black line represents the one-to-one relation. Printed in the bottom right corners of the panels are the median offsets between the trends and the one-to-one relation, and the standard deviations of the trends. Printed in the top left corners of the panels are the median error bars of the recovered values, defined as the median difference between the 16\textsuperscript{th} and 84\textsuperscript{th} percentiles of each recovered value.}
    \label{fig: Params 2}
\end{figure*}

The good agreement between the observed and model spectra indicates that the recovered parameters of \magphys\ are well constrained; we now investigate their accuracy by comparing them to the known values. A summary of the parameters we investigated and the values of their recovered offsets with the true values are presented in Table~\ref{tab: parameter summary}. These offsets are depicted in Fig.~\ref{fig: Params 1} and Fig.~\ref{fig: Params 2}, and discussed in more detail in the following sections. 

\subsection{Idealised \eagle\ sample}

\subsubsection{Stellar Masses}
We compare the \magphys\ likelihood estimates of stellar mass with the true values for all \eagle\ galaxies at $z=0.10$, $z=2.01$, and $z=5.04$ in panels (a), (b) and (c) of Fig.~\ref{fig: Params 1}, respectively, with each galaxy coloured by the \magphys\ likelihood estimate of average stellar age. We note that there is a clear lower bound to the \eagle\ stellar masses at $\log(\mstar/\msun)=8.5$ as \citet{Camps18} only produced the SEDs of galaxies with stellar masses larger than this value. While the estimated and true stellar masses are well correlated, panel (a) shows that there is a systematic underestimation of stellar mass of $0.140$~dex on average by \magphys\ for the redshift $z=0.10$ subsample, and that at the highest masses, this offset is increased. This systematic offset is within the median error bar for this subsample, however we investigate its potential causes in Sect.~\ref{sect: Discussion}.

Panels (b) and (c) of Fig.~\ref{fig: Params 1} show that the median offset in recovered stellar masses decreases for the higher-redshift subsamples, reaching $0.050$ and $0.070$ dex for the $z=2.01$ and $z=5.04$ samples, respectively. However, the NMAD and standard deviation indicate similar or larger scatter, implying that while the bias decreases, the overall dispersion in the recovered masses does not decrease. We find that at these redshifts the galaxies with younger \magphys\ estimated average stellar ages have higher offsets in stellar mass, while galaxies with older stellar ages are recovered much better. We discuss the impact of average stellar ages on the recovery of stellar masses in Sect.~\ref{sect: Discussion}.

\subsubsection{Star Formation Rates}
Panels (d), (e) and (f) of Fig.~\ref{fig: Params 1} compare the star formation rates provided by the \eagle\ database with the \magphys\ likelihood estimates, again coloured by the \magphys\ average stellar ages. We note that the \eagle\ SFRs are instantaneous, and computed as the sums of the star formation rates of all gas particles within the galaxy at the time the galaxy is recorded; in contrast, the \magphys\ SFRs are averaged over the last 100 Myr. We find that \magphys\ recovers the SFRs of the galaxies well, with low median offset values for all three redshift subsamples, which is likely due to the fact that the mock SEDs are extremely well-sampled from the UV to the submillimetre, and that realistic SFHs are assumed by \magphys\ \citep{daCunha15, Pacifici23}.

In all three redshift subsamples, there is a large degree of scatter evident in the plots. In the redshift $z=2.01$ and $z=5.04$ samples, the younger galaxies show more disagreement than older galaxies, though there is still considerable scatter found among the older galaxies, especially in the redshift $z=0.10$ and $z=2.01$ subsample. Overall, however, the median systematic offsets are low and smaller than their respective median error bars which indicates that \magphys\ is recovering SFR well.

\subsubsection{Specific Star Formation Rates}
The \eagle\ database does not provide the specific star formation rates of galaxies, therefore we calculated the `true' sSFR values by dividing the SFR of galaxies by their stellar masses. The comparisons between these values and the \magphys\ likelihood estimates are presented in panels (g), (h) and (i) of Fig.~\ref{fig: Params 1}. All three redshift subsamples show a strong correlation between sSFR and the average stellar ages of galaxies, with high sSFR galaxies being younger and lower sSFR galaxies being older.

The primary cause of the large scatter across the three redshift subsamples is likely inherited from the scatters of the stellar masses and SFRs, and the systematic uncertainties affecting these two parameters will most likely be the primary influence on the sSFR uncertainties as well. Even with these uncertainties included, the small median offset values lead us to conclude that \magphys\ can recover sSFR to a satisfactory degree.

\subsubsection{Stellar Ages}
The comparisons between the \eagle\ values and the \magphys\ likelihood estimates of mass-weighted stellar ages are presented in panels (a), (b), and (c) of Fig.~\ref{fig: Params 2}, and coloured by the \magphys\ likelihood estimates of stellar mass.
We find that there is a severe underestimation of a majority of the average stellar ages by \magphys. For the redshift $z=0.10$ subsample, the median offset value of $0.261$ dex and the median error bar value of $0.345$ dex are the largest of the six parameters we investigated at any redshift. In the redshift $z=2.01$ and $5.04$ subsamples, we find similar severe underestimations by \magphys, despite the relatively low median offset values of $0.096$ and $0.112$, respectively. In general, \magphys\ can underestimate the stellar mass-weighted ages by up to a factor of 10 in the worst cases, though we also note the very small spread of the true values from the \eagle\ galaxies. Due to the known difficulties of recovering the SFH of galaxies using only photometric methods (e.g., \citealt{Carnall19, Leja19a}), we investigated the cause of these underestimations and their impact on other parameters, specifically the recovered stellar masses, in Sect.~\ref{sect: Discussion}.

\subsubsection{Dust Luminosities}
The \eagle\ database does not provide the dust luminosities of its galaxies, therefore we calculated the `true' value of \ldust\ for each galaxy by directly integrating the SEDs computed by \skirt\ between 8 and 1000 $\mu$m. We note that \magphys\ defines \ldust\ by integrating the dust emission SED between 3 and 1000 $\mu$m. We could not integrate the \skirt\ SEDs in that exact range as they do not separate the dust and stellar emission, and the mid-IR continuum between 3 and 8 $\mu$m is severely contaminated by stellar emission. However, this should only introduce differences of at most a few percent in \ldust. Panels (d), (e) and (f) of Fig.~\ref{fig: Params 2} show the comparisons between these `true' values and the \magphys\ likelihood estimates of dust luminosity, coloured by the \magphys\ median-likelihood estimates of average stellar ages.

The dust luminosity is the parameter derived with the most accuracy in our analysis. This is very likely thanks to the very well-sampled input SEDs, especially in the far-IR region where the dust emission peaks. In the redshift $z=5.04$ subsample, there is a visually obvious correlation between dust luminosity and mass-weighted stellar age, with the brightest dust luminosity galaxies being the youngest. Across all three redshift subsamples, the galaxies with older mass-weighted stellar ages have the highest scatter, however the overall scatter of all three redshift subsamples is very low. As the median offset values of all three redshift subsamples is lower than $\sim$$0.07$ dex, we conclude that \magphys\ recovers dust luminosities very well.

\subsubsection{Dust Masses}
The comparisons between the \eagle\ dust mass values and the \magphys\ likelihood estimates of dust mass are presented in panels (g), (h) and (i) of Fig.~\ref{fig: Params 2}, and coloured by the \magphys\ median-likelihood estimates of average stellar age. To compare these sets of values, we accounted for the difference in dust models used by \citet{Camps18} for the \eagle\ database, and by the \magphys\ model. \cite{Camps18} used the \citep{Zubko04} dust model with an absorption coefficient at 350 $\mu$m of $\kappa_{350} = 0.330$ cm$^{2}$ kg$^{-1}$ to calculate the dust properties of the \eagle\ galaxies, while \magphys\ assumes the \citet{Dunne00} value for an absorption coefficient at 850 $\mu$m of $\kappa_{850} = 0.077$ cm$^{2}$ kg$^{-1}$. Therefore, we multiplied the recovered \magphys\ dust mass values by a factor of $1.376$ for all of the three redshift subsamples before comparing our results. Once that correction is made, we find that \magphys\ recovers the dust masses of \eagle\ galaxies very accurately.

We note that \citet{Hayward15} found a systematic underestimation of $\sim0.2-0.3$ dex in their recovered \magphys\ dust mass values. They attribute this to the assumptions of the cold dust phase of the ISM made by the simulations they investigated. We found no such underestimation in our results, with the largest median offset of our redshift subsamples being $0.071$ dex, which occurs for the redshift $z=0.10$ subsample. In the redshift $z=5.04$ subsample, we find a strong correlation between dust mass and average stellar age, and find that the galaxies with younger stellar ages tend to have their higher dust masses overestimated slightly by \magphys, however this trend is not present in the other redshift subsamples. Overall, the agreement is very good, and again it can be attributed to the good IR SED sampling of the mock galaxies.

\subsection{WAVES-like sample}

\begin{figure*}
    \centering
    \includegraphics[width=0.8\linewidth]{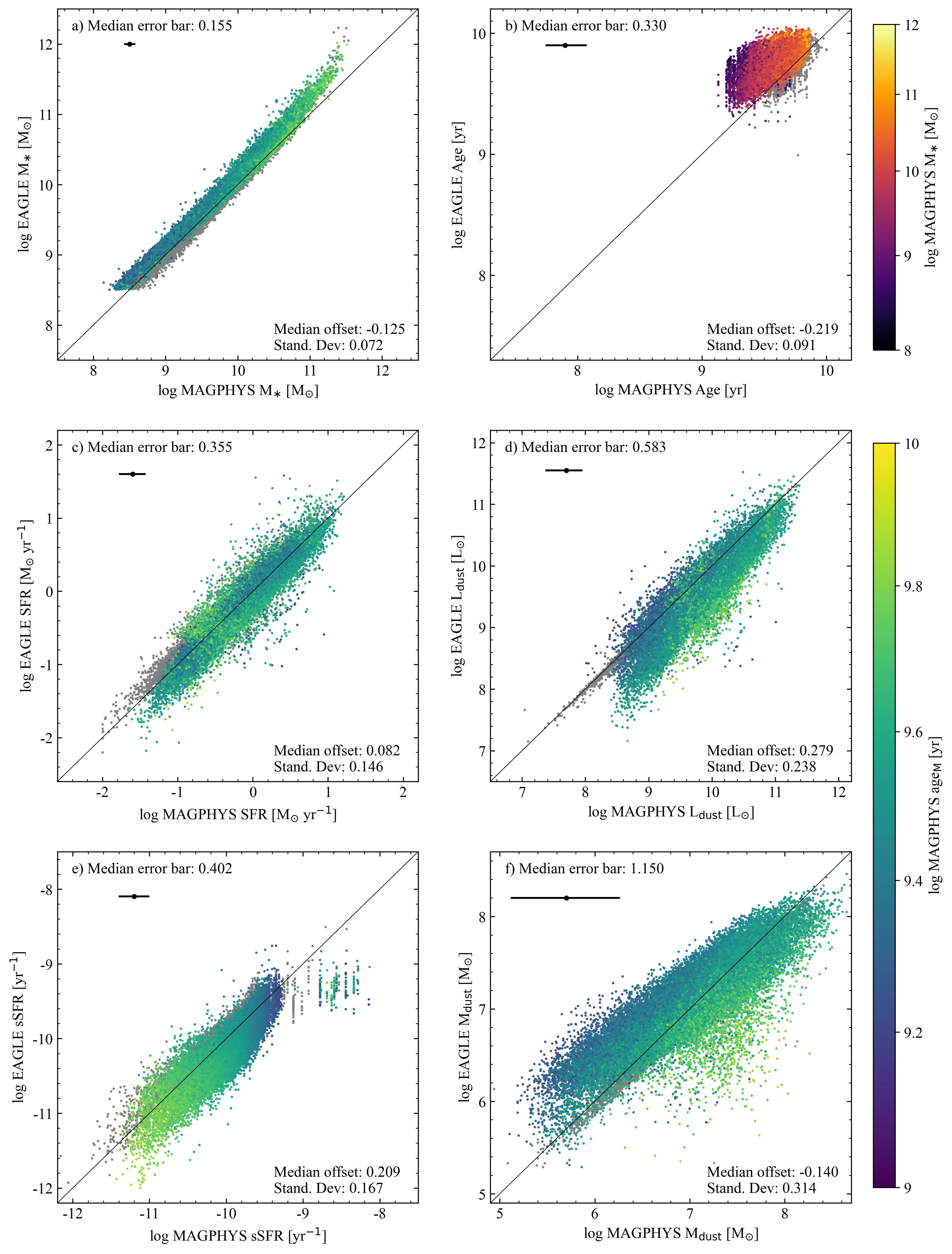}
    \caption[...]{Comparisons between the true parameters of the \eagle\ mock galaxies ($y$-axes), and the median-likelihood parameters recovered by \magphys\ from fitting their \skirt-processed SEDs for a WAVES-like filter set ($x$-axes). The panels show the comparisons of a) stellar masses, b) average stellar ages, c) SFR, d) dust luminosities, e) sSFR, and f) dust masses. The parameters are coloured by recovered \magphys\ average stellar age, except for average stellar age which is coloured by recovered \magphys\ stellar mass. For ease of comparison, the grey dots show the sample of redshift $z=0.10$ galaxies fit using the adjusted age prior, as seen in Fig.~\ref{fig: OA Age and M*} and Fig.~\ref{fig: All Old Ages}. The diagonal black lines represent the one-to-one relation. Printed in the bottom right corners of the panels are the median offsets between the trends and the one-to-one relation, and the standard deviations of the trends. Printed in the top left corners of the panels are the median error bars of the recovered values, defined as the median difference between the 16\textsuperscript{th} and 84\textsuperscript{th} percentiles of each recovered value.}
    \label{fig: WAVES params}
\end{figure*}

\begin{table*}
    \centering
    \resizebox{\textwidth}{!}{
    \begin{tabular}{| l | l | rrrrrrr |}
        \hline
        Parameter & \eagle\ Database Name & Median Offset & Standard Deviation & NMAD & Offset Q1 & Offset Q3 & Outlier Fraction & Median Error Bar \\
        \hline
        \multicolumn{9}{|c|}{Adjusted Age Prior ($z=0.10$) Subsample} \\
        \hline
        $\log(\mstar/\msun)$ & MassType\_Star & -0.029 & 0.083 & 0.065 & -0.073 & 0.015 & 0.042 & 0.140 \\
        $\log(\rm{SFR}/\msun\,\rm{yr}^{-1})$ & StarFormationRate & -0.071 & 0.112 & 0.086 & -0.132 & -0.015 & 0.040 & 0.105 \\
        $\log(\rm{sSFR}/\rm{yr}^{-1})$ & - & -0.047 & 0.134 & 0.115 & -0.125 & 0.031 & 0.028 & 0.207 \\
        $\log(\agem/\rm{yr})$ & InitialMassWeightedStellarAge & -0.081 & 0.073 & 0.068 & -0.126 & -0.034 & 0.018 & 0.225 \\
        $\log(\ldust/L_\odot)$ & - & 0.011 & 0.042 & 0.033 & -0.011 & 0.034 & 0.025 & 0.070 \\
        $\log(\mdust/\msun)$ & Mass\_Dust & 0.069 & 0.058 & 0.049 & 0.034 & 0.101 & 0.022 & 0.085 \\
        \hline
        \multicolumn{9}{|c|}{WAVES-like ($z \leq 0.27$) Subsample} \\
        \hline
        $\log(\mstar/\msun)$ & MassType\_Star & -0.125 & 0.072 & 0.066 & -0.167 & -0.078 & 0.017 & 0.155 \\
        $\log(\rm{SFR}/\msun\,\rm{yr}^{-1})$ & StarFormationRate & 0.082 & 0.146 & 0.118 & -0.001 & 0.158 & 0.038 & 0.355 \\
        $\log(\rm{sSFR}/\rm{yr}^{-1})$ & - & 0.209 & 0.167 & 0.141 & 0.106 & 0.298 & 0.033 & 0.402 \\
        $\log(\agem/\rm{yr})$ & InitialMassWeightedStellarAge & -0.219 & 0.091 & 0.085 & -0.277 & -0.163 & 0.017 & 0.330 \\
        $\log(\ldust/L_\odot)$ & - & 0.279 & 0.238 & 0.211 & 0.139 & 0.424 & 0.027 & 0.583 \\
        $\log(\mdust/\msun)$ & Mass\_Dust & -0.140 & 0.314 & 0.293 & -0.325 & 0.073 & 0.021 & 1.150 \\
        \hline
    \end{tabular}}
    \caption[...]{The six parameters we investigated to determine the accuracy of the \magphys\ model for our two additional samples discussed in Section~\ref{sect: Discussion}: the adjusted age prior sample at redshift $z = 0.10$, and the WAVES-like sample at redshift $z \leq 0.27$. The columns show parameter names and units, parameter names used in the \eagle\ database (if they are provided), median offsets, standard deviations, NMAD, Q1 offsets, Q3 offsets, outlier fractions, and median error bars of the recovered \magphys\ values when compared to the true \eagle\ values.}
    \label{tab: parameter summary 2}
\end{table*}

Here we compare our results for the more realistic, WAVES-like sample of mock galaxies. The comparisons between the true \eagle\ parameters and the parameters recovered by \magphys\ are shown in Fig.~\ref{fig: WAVES params}, with the corresponding statistical metrics listed in Table~\ref{tab: parameter summary 2}. As for the idealised samples, we apply the same emissivity correction factor to the dust masses. In addition, we adopt a restricted age prior (discussed in Sect.~\ref{sect: Discussion}) in order to minimise systematic stellar-mass offsets associated with mismatches between the \magphys\ SFH priors and the \eagle\ age distribution.

From Fig.~\ref{fig: WAVES params} and Table~\ref{tab: parameter summary 2}, we find that stellar masses and star formation rates are the most robustly recovered parameters in the WAVES-like sample, with median offsets of $-0.125$ and $0.082$~dex, respectively. This result is expected given that the nine photometric bands used in this analysis provide continuous coverage of the optical and near-infrared SED, which strongly constrains the integrated stellar emission and dust attenuation. This wavelength range captures contributions from both young and evolved stellar populations and helps to partially mitigate age-dust degeneracies, even in the absence of ultraviolet or infrared data.

In contrast, dust luminosities and dust masses are substantially less well constrained. These parameters exhibit large median offsets, increased scatter, and broad posterior uncertainties, reflecting the lack of mid- and far-infrared photometry that directly traces dust emission. While energy balance allows the optical-to-near-infrared SED to place weak constraints on the total dust luminosity (e.g. \citealt{magphys, daCunha13b}), these constraints are necessarily prior-dominated in the absence of infrared data. Dust mass estimates are even more uncertain, as they depend additionally on assumptions about dust temperature and emissivity (e.g. \citealt{daCunha21}). The WAVES-like results therefore illustrate the limited ability of broad-band optical-near-infrared photometry alone to recover dust-related parameters with high accuracy.

The increased uncertainty in the dust properties has secondary effects on other parameters. In particular, less well-constrained dust attenuation leads to enhanced age-dust degeneracies, which contribute to the modest systematic offsets observed in stellar masses and star formation rates for this sample. In this regime, dust luminosity is underestimated by $\sim$0.28~dex on average, consistent with the lack of infrared constraints, and this propagates into the recovered stellar population properties.

Specific star formation rates and mass-weighted stellar ages are also recovered with larger systematic offsets than in the idealised case, although their scatter and posterior uncertainties remain smaller than those of the dust parameters. The sSFR offset largely reflects the stellar-mass offset, while the challenges associated with recovering reliable stellar ages from broad-band photometry are discussed in detail in Sect.~\ref{sect: Discussion}.

As the primary science goals of the WAVES survey focus on measuring stellar masses, star formation rates, and halo properties in the low-redshift Universe, the reduced accuracy of dust-related parameters in this WAVES-like analysis is not unexpected and does not compromise the survey's core objectives. Instead, these results provide a clear demonstration of how limited wavelength coverage affects parameter recovery, and highlight the need for caution when interpreting dust properties, or stellar ages that are strongly coupled to dust attenuation, in the absence of infrared data (see also, e.g., discussion in \citealt{Li24}).

\section{Discussion} \label{sect: Discussion}

\subsection{What causes the stellar mass offsets?}

\begin{figure}
    \centering
    \includegraphics[width=\columnwidth]{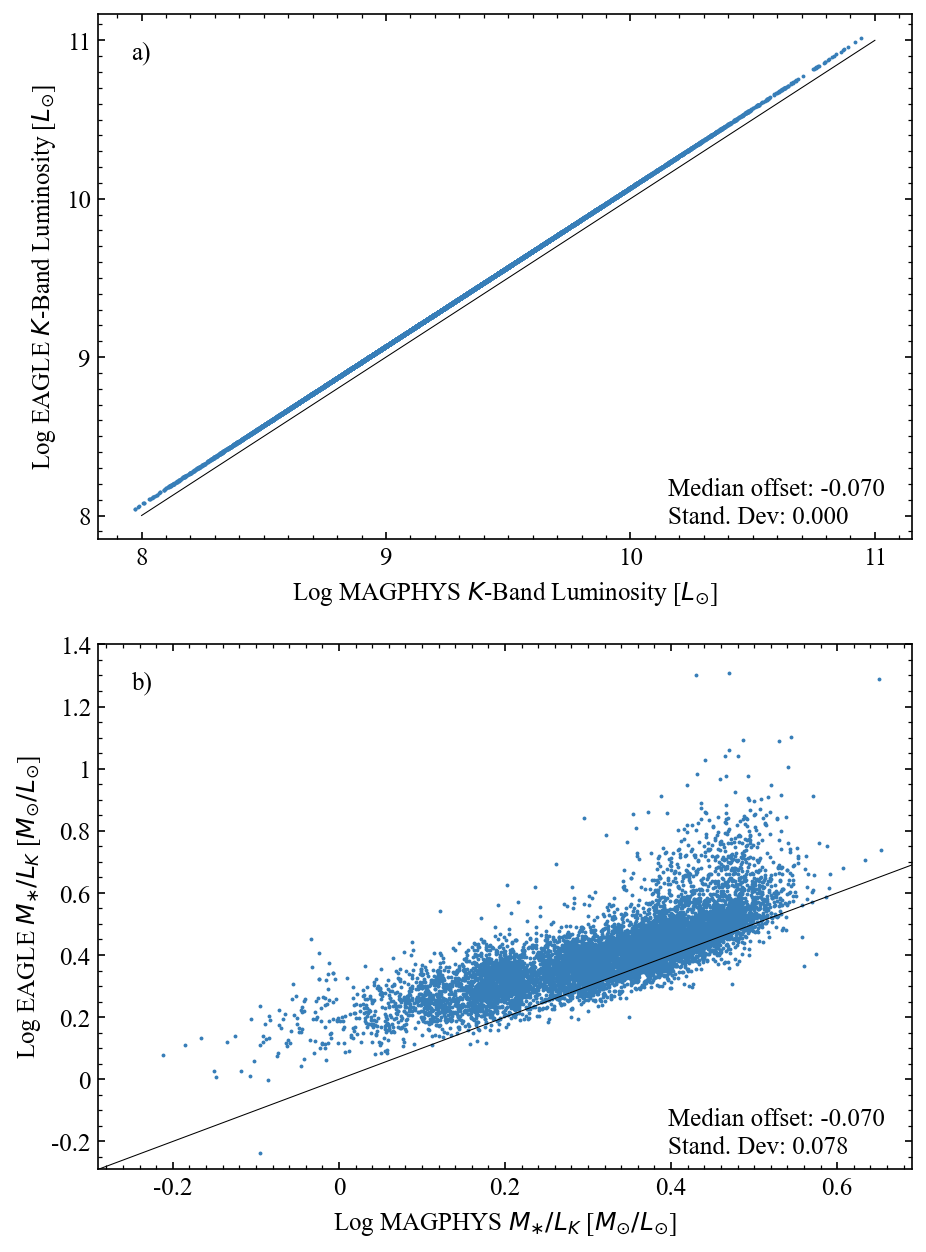}
    \caption[The $K$-band luminosities and mass-to-light ratios of the redshift $z=0.10$]{The UKIDSS $K$-band luminosities and the mass-to-light ratios of 10,288 \eagle\ galaxies at redshift $z=0.10$, displayed in panels a) and b), respectively. The mass-to-light ratios are specifically the ratios of stellar mass in M\textsubscript{$\odot$} to $K$-band luminosity in L\textsubscript{$\odot$}. The true \eagle\ values are on the y-axes and the recovered \magphys\ values are on the x-axes. The diagonal black line represents the one-to-one relation. Printed in the bottom right corner is the median offset between the trend and the one-to-one relation, and the standard deviation of the trend.}
    \label{fig: Kband and ratio}
\end{figure}

The results in Sect.~\ref{sect: Results} show that most parameters are recovered well by \magphys. However, there is a significant systematic offset of about $0.140$ dex in the recovered mass-weighted stellar ages of the galaxies in the redshift $z=0.10$ subsample, which may be related to offsets in the recovered stellar ages. Given that stellar mass is one of the fundamental parameters of galaxies that we expect to recover from SED fitting, this warrants further investigation. Both \eagle+\skirt\ and \magphys\ use the stellar population synthesis models of \cite{Bruzual03} to model the stellar emission in galaxies, and they both assume the \cite{Chabrier03} initial mass function. Therefore, differences in modelling stellar populations, or in the assumed IMF, cannot be the cause of this offset. 

The stellar mass of a galaxy is dominated by low-mass stars which emit primarily in the near-infrared. We investigated whether \magphys\ was accurately modelling the near-IR emission of the redshift $z=0.10$ subsample galaxies, and found a slight systematic offset ($-0.07$~dex) between the $K$-band luminosities provided by the \eagle\ database and the $K$-band luminosities of the best-fit \magphys\ model, shown in Fig.~\ref{fig: Kband and ratio} (a). \magphys\ underestimates the $K$-band luminosities by $0.07$~dex, and we note that this offset is very consistent across the sample. This translates into a similar offset in $K$-band mass-to-light ratio. Given that our observed stellar mass offset is even larger (Fig.~\ref{fig: Params 1}), this offset is not the sole contributor to the difference in stellar mass.

The mass-to-light ratios of a galaxy in the near-IR are affected by its star formation history (SFH) and secondarily affected by its dust content (e.g., \citealt{Chaves-Montero20}). The star formation history of a galaxy describes the star formation rate of the galaxy over the length of its existence, including any bursts of star formation due to interactions with other galaxies or similar processes. It is notoriously difficult to recover the SFH of a galaxy from its broad-band SED alone, as multiple different SFH can result in similar looking SEDs \citep{Nersesian23, Nersesian25, Csizi24}, and this is further compounded by age-dust degeneracies that affect the UV/optical colours. Spectral energy distribution models use various templates or parametric functions that attempt to replicate the complex history of star formation in a galaxy over its lifetime. For example, \magphys\ uses a linearly rising and then exponentially declining (`delayed-tau model') with superimposed random bursts to generate its library of possible SFHs for input SEDs (see section 3.1.1 of \citealt{daCunha15}). A difference in the simulated star formation histories of the \eagle\ galaxies and the possible templates generated by \magphys\ could be indicated by the significant discrepancies in the stellar ages, and is a plausible cause of the difference in the current stellar masses \citep{Carnall19, Leja19a}.

\begin{figure}
    \centering
    \includegraphics[width=\columnwidth]{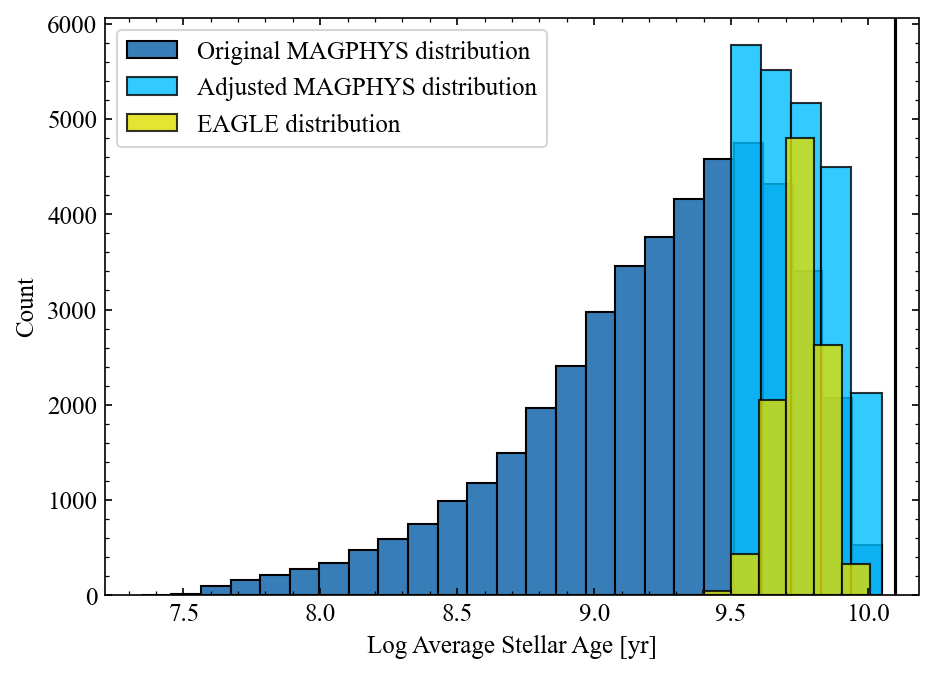}
    \caption[Distributions of the original \magphys\ mass-weighted age prior, adjusted \magphys\ age prior, and \eagle\ age values]{The distributions of the original \magphys\ age prior (blue), the adjusted \magphys\ age prior (light blue), and the age values of all 10,288 \eagle\ galaxies (yellow) from the redshift $z=0.10$ subsample of galaxies. The vertical black line at 10\textsuperscript{10.1} years represents the oldest age a stellar population can be at this redshift (set by the age of the Universe).}
    \label{fig: Age Hist}
\end{figure}

\begin{figure}
    \centering
    \includegraphics[width=\columnwidth]{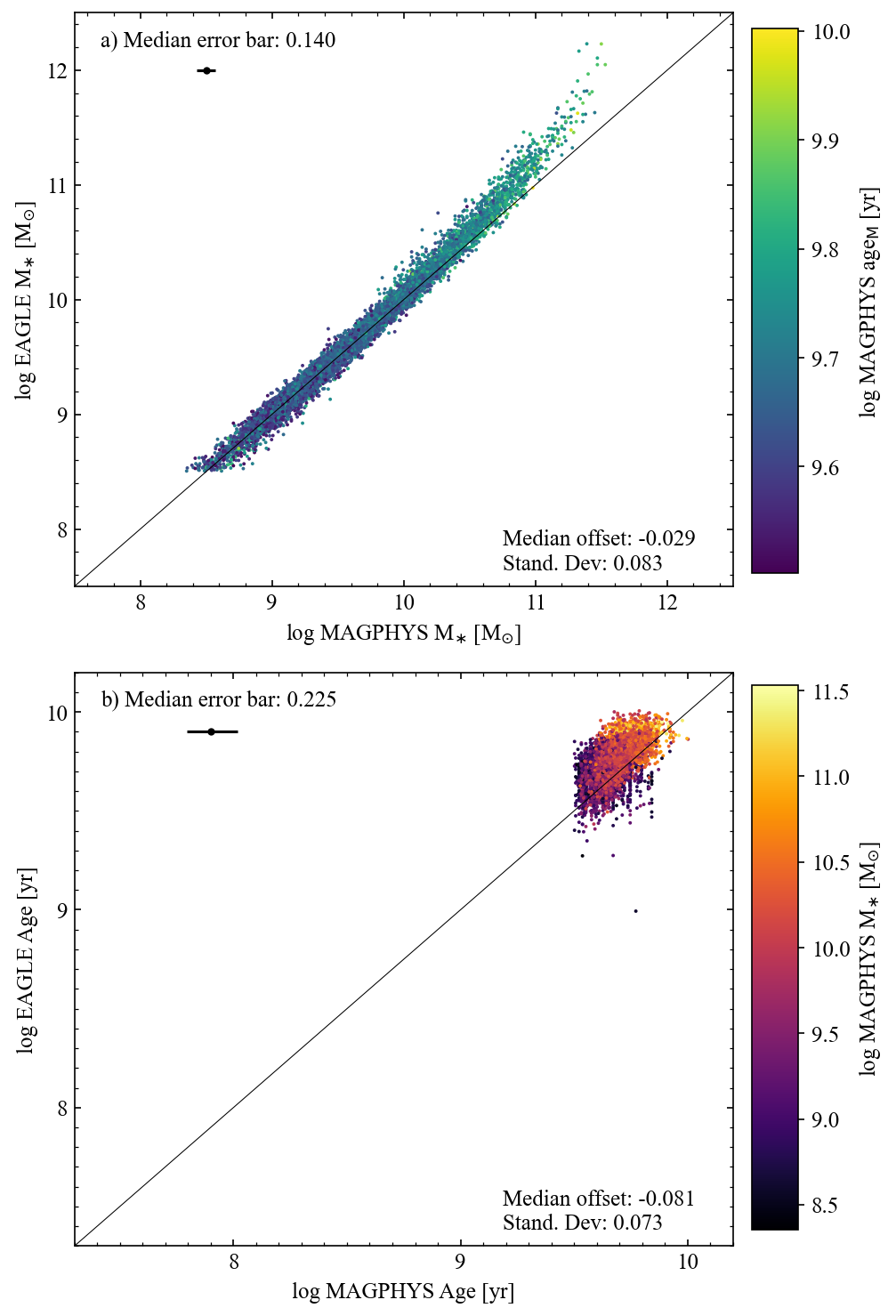}
    \caption[Comparison of stellar ages and stellar mass, with adjusted age prior]{Comparisons of the stellar masses and mass-weighted stellar ages of the 10,288 \eagle\ galaxies from the redshift $z=0.10$ subsample, after removing younger ages from the \magphys\ age distribution, in panels a) and b), respectively. The diagonal black line represents the one-to-one relation. Printed in the bottom right corners of the panels are the median offsets between the trends and the one-to-one relation, and the standard deviations of the trends. Printed in the top left corners of the panels are the median error bars of the recovered values, defined as the median difference between the 16\textsuperscript{th} and 84\textsuperscript{th} percentiles of each recovered value.}
    \label{fig: OA Age and M*}
\end{figure}

In Fig.~\ref{fig: Age Hist}, we compare the distribution of the \magphys\ mass-weighted age priors with the distribution of all the age values from the \eagle\ models for the galaxies in our redshift $z=0.10$ subsample. The \eagle\ mass-weighted stellar ages range almost exclusively from $10^{9.5}$ to $10^{10}$ years. This is a very small and old section of the range covered by the \magphys\ age priors, which extend to mass-weighted ages as low as $\sim$$10^{7.5}$ years. To investigate the effect of this discrepancy on the recovered parameters, we generated a new \magphys\ age prior adjusted to match the \eagle\ age distribution, which is shown in light blue in Fig.~\ref{fig: Age Hist}. Essentially, we exclude \magphys\ models with ages younger than $10^{9.5}$ years from the prior distribution. We then re-fit the redshift $z=0.10$ subsample with this new age prior and compare the recovered stellar masses and mass-weighted ages with their true values in Fig.~\ref{fig: OA Age and M*}. We note that the other galaxy parameters (SFR, sSFR, \mdust, \ldust) remain well-constrained with low offsets, and for completeness we show the new comparisons for these parameters in Appendix~\ref{App: All Old Ages APP}. The statistics of these comparisons are also displayed in Table~\ref{tab: parameter summary 2}.

Fig.~\ref{fig: OA Age and M*} shows that we recover the stellar masses of the \eagle\ mock galaxies with much smaller offsets when using the adjusted \magphys\ age prior. The median offset value of the stellar mass offset has been reduced from $0.140$ dex in Fig.~\ref{fig: Params 1} (a) to $\mathbf{0.029}$ dex in Fig.~\ref{fig: OA Age and M*} (a). A similar reduction is found in the mass-weighted stellar ages, with the median offset value changing from $0.261$ dex in Fig.~\ref{fig: Params 2} (a) to $\mathbf{0.081}$ dex in Fig.~\ref{fig: OA Age and M*} (b).

These results support our hypothesis that the offsets of the stellar ages and the stellar masses found when recovering the physical parameters of \eagle\ galaxies with \magphys\ are caused by the differences between the histories of the \eagle\ galaxies and the SFH prior of the \magphys\ model. It is interesting that even though \magphys\ is able to fit the SEDs of the \eagle\ galaxies well, the choice of SFH prior has a significant effect on the recovery of the stellar masses of those galaxies. We note that similar results were found by \cite{Pforr12} in a test fitting the UV-to-near-IR SEDs of $z\sim2$ mock galaxies from the semi-analytic model. They concluded that the stellar masses of galaxies underestimated because the mock galaxy SFHs were rising, whereas their SED model assumed an exponentially-declining form. When using rising SFHs, the offsets in stellar masses decreased.

Our results beg the question: are the \magphys\ SFH priors inappropriate to model low-redshift galaxies? While they are certainly `too young' for the subset of \eagle\ redshift $z=0.10$ galaxies presented here, we must be cautious about taking these results as a sign that the \magphys\ priors need fine-tuning. This is because there are indications that the \eagle\ stellar ages may not be representative of the stellar ages of galaxies in the local Universe. For example,  \citet{vandesande19} compared galaxies at $z \simeq 0$ from multiple simulations, including \eagle, to galaxies from integral field unit (IFU) surveys such as the SAMI (Sydney-AAO Multi-object Integrated field spectrography) Galaxy Survey \citep{SAMI12, SAMI15}, the ATLAS\textsuperscript{3D} Survey \citep{ATLAS11}, the SAURON Survey \citep{SAURON02}, the CALIFA (Calar Alto Legacy Integral Field Area) Survey \citep{CALIFA12}, and the MASSIVE Survey \citep{MASSIVE14}. They find that within the same stellar mass range ($\log(\mstar/\msun)<11.5$) the simulated \eagle\ galaxies tend to have much older luminosity-weighted ages than the observed galaxies of their sample. They find no \eagle\ galaxies with ages $<10^{9.4}$ yr, while the sample of observed galaxies extends to ages below $10^{8.6}$ yr.

\begin{figure}
    \centering
    \includegraphics[width=\linewidth]{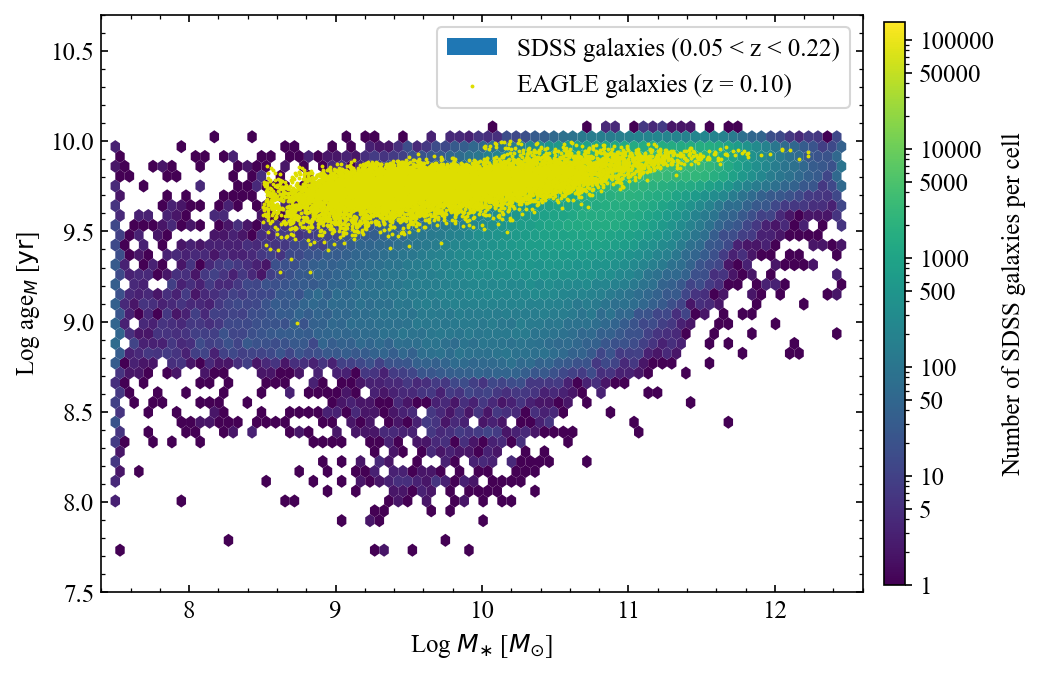}
    \caption[Comparison of \eagle\ and SDSS data]{Comparison of the stellar age distribution of \eagle\ and SDSS DR4 galaxies. The \eagle\ galaxies are shown in yellow, and the SDSS galaxies are coloured by the number of galaxies per cell as seen in the colour bar. The stellar mass lower limit of the \eagle\ galaxies at 10\textsuperscript{9.5} M\textsubscript{$\odot$} is clear when compared to the lower limit of the SDSS galaxies at 10\textsuperscript{7.5} M\textsubscript{$\odot$}.}
    \label{fig: SDSS}
\end{figure}

We further investigate how the average stellar ages of our \eagle\ galaxies compared to those of observed galaxies at redshift $z=0.10$ by analysing data from the Sloan Digital Sky Survey (SDSS). We compared the distribution of average stellar ages of 567,486 SDSS galaxies that lie between $0.005<z<0.22$, with a median redshift of $z = 0.13$, from the fourth SDSS data release (DR4; \citealt{Gallazzi05, Gallazzi06}), with the distribution of the \eagle\ galaxies at redshift $z=0.10$. Fig.~\ref{fig: SDSS} shows the age-mass relation of the SDSS galaxies and the \eagle\ galaxies. The SDSS average stellar ages are weighted by luminosity rather than mass, so the two ages are not directly comparable: mass-weighted ages tend to be older since the luminosity is more dominated by younger stars than the mass. For reference, the difference between mass- and $r$-band luminosity-weighted ages for the broad range of SFHs in the \magphys\ model library is $\log(\mathrm{age}_M/\mathrm{age}_r)=0.18\pm0.14$. Even when accounting for this in Fig.~\ref{fig: SDSS}, we can see that the \eagle\ galaxies only occupy a small part of the full range of the average stellar ages of the SDSS galaxies, and are missing many of the younger galaxies found in the SDSS.

We conclude that the \eagle\ galaxies we used may not be representative of the full range of average stellar ages and SFH in the local Universe, and therefore a generalised adjustment of the \magphys\ priors is not necessarily required at this point. However, these results highlight the complexity of recovering even basic parameters such as the stellar masses of galaxies. While the SED fits were of good quality with relatively small residuals, variations in the assumptions of the SFH means that a significant amount of stellar mass can be `hidden' without affecting the observed SEDs, as the near-IR SED of old stellar populations does not evolve significantly after a few billion years. Therefore, the broad-band SEDs have little constraining power on the SFHs, especially after a given age \citep{Hayward15, Chaves-Montero20, Csizi24}. This has been pointed out by other studies. A recent study by \citet{Nersesian23} suggests that the photometry-based estimations of the properties of stellar populations, including their average stellar ages, are heavily influenced by the modelling approach and therefore the assumptions made by SED models. It is likely then that an accurate star formation history, and therefore average stellar age, cannot be recovered from broad-band SED modelling alone. The addition of spectroscopic analysis is also required to accurately constrain stellar ages via absorption features \citep{Carnall19, Leja19a, Leja19b, Nersesian25}.

During our investigation, we found that the \magphys\ model is more accurate when recovering the stellar mass values of higher redshift galaxies. The stellar mass underestimation present in the redshift $z=0.10$ galaxies is reduced at redshift $z=2.01$ and redshift $z=5.04$, as seen in panels (a), (b) and (c) of Fig.~\ref{fig: Params 1}. These reduced stellar mass offsets are likely due to the average stellar age distributions being more similar to the \eagle\ distributions at these higher redshifts. Additionally, at higher redshifts, the range of possible SFHs and therefore average stellar ages is reduced due to there being less time since the formation of the galaxies, and therefore the derived galaxy properties are less sensitive to the SFH prior. This is more evidence that the underestimation by \magphys\ when recovering the stellar masses of redshift $z=0.10$ galaxies is due to the difference in the distributions of the average stellar ages of the galaxies and the SFH prior, and that when the distributions of the \magphys\ age priors better match the \eagle\ age distributions, this underestimation is reduced.

\subsection{Methodological limitations and scope of this study}

As with any validation study based on simulated galaxies, the results presented here are subject to limitations arising from the underlying simulations and modelling assumptions. It is therefore important to clarify the scope of this work and the context in which our conclusions should be interpreted.

First, our analysis relies on galaxies from the \eagle\ simulations, post-processed with the \skirt\ radiative transfer code. While this provides a powerful and self-consistent framework in which the intrinsic physical parameters of galaxies are known, it necessarily inherits assumptions related to subgrid physics, dust prescriptions, and radiative transfer modelling, including finite resolution effects (e.g., \citealt{Hayward15}). Our aim is not to claim that \eagle+\skirt\ provides a complete or fully representative description of the observed galaxy population, but rather to use it as a controlled testbed that enables a direct assessment of the accuracy of SED fitting under known conditions. This controlled setting allows us to isolate the impact of SED modelling assumptions, particularly those related to star formation histories and priors, in a way that is not possible using observational data alone.

Second, part of our analysis intentionally adopts idealised conditions, using well-sampled UV-to-submillimetre SEDs. This choice is motivated by the need to establish an upper bound on the performance of the \magphys\ model and to identify intrinsic limitations of the SED fitting approach when observational constraints are maximised. Such conditions are rarely achieved in practice, therefore we complement this analysis with a WAVES-like mock sample that employs limited optical–near-infrared wavelength coverage and survey-level photometric uncertainties. This additional test demonstrates how parameter recovery degrades under more realistic observational conditions and clarifies which parameters remain robust in survey applications.

Third, we restrict our analysis to star-forming galaxies and exclude systems dominated by active galactic nuclei. AGN emission can significantly alter galaxy SEDs, particularly in the UV and mid-infrared, and requires dedicated modelling components that are not fully included in the \magphys\ framework. While assessing the impact of AGN contamination on SED fitting performance is an important observational challenge, it lies beyond the scope of the present study and would require a different modelling approach (e.g., \citealt{Yang20}).

Finally, the simulated galaxy sample explored here does not span the full diversity of stellar ages, metallicities, and star formation histories observed in real galaxies, particularly at low redshift. As discussed above, this limitation is central to one of our key results: mismatches between the intrinsic SFHs of the simulated galaxies and the priors adopted in SED fitting can lead to systematic biases in recovered stellar masses, even when the SEDs are well fitted. We therefore caution against interpreting the specific magnitude of the stellar-mass offsets reported here as universally applicable. Instead, the broader implication is that SED-derived stellar masses and ages are intrinsically sensitive to SFH assumptions, and that this sensitivity should be accounted for when interpreting results from both simulated and observed galaxy samples.

\section{Summary \& Conclusion} \label{sect: Conclusion}

In this paper, we tested the ability of the \magphys\ SED model \citep{magphys, daCunha15} to recover accurate galaxy parameters from the observed emission of galaxies. To do this, we used simulated galaxies from the \eagle\ simulations, which have known physical parameters, and for which the UV-to-IR emission in several photometric bands was computed using the \skirt\ radiative transfer code and published by \cite{Camps18}. We used \magphys\ to fit the SEDs, and compared the physical parameters constrained by \magphys\ with the true parameters provided by the \eagle+\skirt\ database. We performed this test on a large, idealised sample of simulated galaxies spanning a range in redshifts: $10,288$ galaxies at $z=0.10$, $18,206$ galaxies at $z=2.01$, and $2,434$ galaxies at $z=5.04$, as well as a more realistic sample of $35,051$ $z\leq0.27$ galaxies that aims to match the properties of the upcoming WAVES-Wide spectrocopic sample.

We found that overall \magphys\ recovers well the stellar masses, dust masses, dust luminosities, star formation rates, and specific star formation rates of the simulated \eagle\ galaxies in the idealised sample at various redshifts. For all five parameters, the median values of the offsets between the true \eagle\ values and the recovered \magphys\ values were lower than or very similar too the median error bars of the recovered values. However, we noticed a systematic offset when recovering the stellar masses at $z=0.10$, and also that the average stellar ages are not recovered well by \magphys\ at the three redshifts. We found that this is due to the significant difference between the distribution of the \magphys\ age prior and the distribution of the true \eagle\ ages, especially in the lowest redshift studied. The \eagle\ ages are very old relative to the \magphys\ age prior distribution, and we hypothesised that \magphys\ was associating the emission of most of these galaxies with younger stellar populations, leading to an underestimation of the stellar mass. To test this, we adjusted the distribution of the \magphys\ age prior to match the \eagle\ distribution more closely and fitted the SEDs again using this new prior. When comparing the physical parameters recovered using this adjusted prior with the true values, we found that the average stellar age offset and the stellar mass offset had both reduced. We therefore conclude that the stellar mass offset was caused primarily by this difference in average stellar ages, and thus a difference in how the \eagle\ simulations and \magphys\ model the star formation histories of galaxies. We note, however, that there are no differences in the goodness-of-fit when using the two different age priors, meaning that our broad-band SED fitting alone cannot distinguish between different average stellar population ages (as also found in \citet{Nersesian23}). We also note that the other physical parameters studied (SFR, sSFR, \mdust, and \ldust) are just as well recovered with either age prior.

These findings would be concerning for the ability of \magphys\ to recover stellar masses, and undoubtedly demonstrate the impact of star formation history priors on the results of SED fitting codes, however, we find that the ages of the \eagle\ galaxies are likely not representative of the ages of real galaxies in the local Universe \citep{vandesande19, Nersesian23}. Therefore, we conclude that the stellar mass offset we recovered is unique to the \eagle\ simulations, and that such an offset is not guaranteed to arise when applying the \magphys\ model to other galaxy samples, whether simulated or observed. We therefore caution against `fixing' the age prior distribution of \magphys\ to remove this offset. This may cause \magphys\ to recover the stellar masses and ages of low-redshift \eagle\ galaxies more accurately, however it will likely also make \magphys\ less accurate when recovering the parameters of real galaxies. This demonstrates how the chosen SFH of a model can have a significant effect when recovering the stellar masses of galaxies, even when their SEDs are very well sampled by the SED model. It is likely the case that accurate SFH cannot be recovered from SED fitting alone.

For the WAVES-like subsample, the stellar masses and star formation rates are well constrained, albeit with larger dispersion due to the sparser sampling of the SEDs and more realistic uncertainties. This demonstrates that survey-quality optical-near-infrared photometry, when combined with SED modelling, is sufficient for robust measurements of galaxy growth, even though dust-related parameters remain poorly constrained in the absence of infrared data.

Our investigation shows that while \magphys\ is able to recover fundamental galaxy parameters with relatively high accuracy, even a simple parameter such as stellar mass can be affected by a disagreement between the real star formation histories of galaxies and the priors assumed in SED fitting with a systematic difference of about 0.14 dex. This is true even in the case of our study, which used fifty photometric bands from the UV to submm wavelengths; real galaxy SEDs are usually not as well sampled. Given that the SFHs are unlikely to be accurately recovered using broad-band photometry alone (e.g., \citealt{Nersesian23}), we advise that a wide range of SFHs are included in SED fitting priors, so that the  uncertainty due to this parameter is included in the uncertainties of the derived physical parameters, specifically of the stellar masses and ages.

Overall, this study shows that while SED fitting can robustly recover key galaxy properties in survey contexts, its limitations are fundamentally driven by star formation history assumptions and wavelength coverage, rather than by the fitting methodology itself.

\section*{Acknowledgements}

We thank the anonymous referee for their detailed and constructive comments, which have helped us significantly improve the clarity and scope of this work. ZRJ thanks Claudia Lagos for her assistance with the \eagle\ data, and Juno Li for his guidance on technical matters. We also thank the \eagle\ and \skirt\ teams for making their data and code publicly available. ZRJ acknowledges funding from the HTM (grant TK202), ETAg (grant PRG1006) and the EU Horizon Europe (EXCOSM, grant No. 101159513). EdC and AB acknowledge support from the Australian Research Council (projects DP240100589 and CE170100013).

ICRAR and the University of Western Australia are located on Whadjuk Noongar land, and the authors acknowledge the Whadjuk Noongar people as the cultural custodians of that land.

\section*{Data Availability}

The \eagle\ database is available here: \url{https://icc.dur.ac.uk/eagle/database.php}. Documentation for the \skirt\ code is available here: \url{https://skirt.ugent.be/root/_home.html}, and the \skirt\ repository is available here: \url{https://github.com/skirt/skirt7}. The \magphys\ code is available from \url{http://www.iap.fr/magphys/}.




\bibliographystyle{mnras}
\bibliography{paper_bib2} 




\appendix

\section{Remaining Recovered Parameters Comparisons} \label{App: All Old Ages APP}

\begin{figure*}
    \centering
    \includegraphics[width=\textwidth]{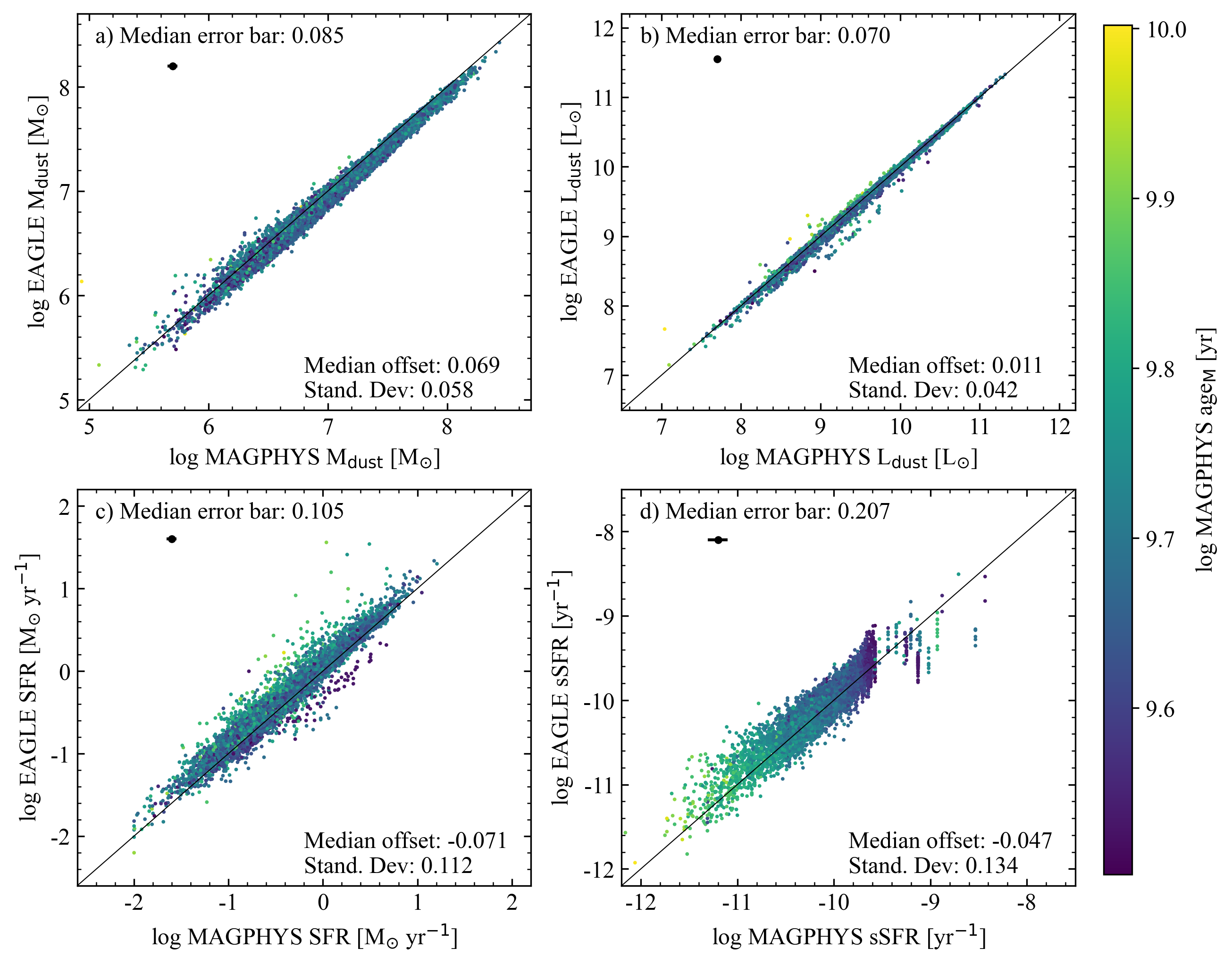}
    \caption{Comparisons between the true parameters of the \eagle\ mock galaxies ($y$-axes), and the median-likelihood parameters recovered by \magphys\ from fitting their \skirt-processed SEDs ($x$-axes). Presented here are the dust masses, dust luminosities, SFR and sSFR of the redshift $z=0.10$ subsample when fitted by \magphys\ with an adjusted age prior (see Section~\ref{sect: Discussion}). All four parameters are coloured by the recovered \magphys\ mass-weighted average stellar ages of the same fitting. The diagonal black line represents the one-to-one relation. Printed in the bottom right corners of the panels are the median offsets between the trends and the one-to-one relation, and the standard deviations of the trends. Printed in the top left corners of the panels are the median error bars of the recovered values, defined as the median difference between the 16\textsuperscript{th} and 84\textsuperscript{th} percentiles of each recovered value.}
    \label{fig: All Old Ages}
\end{figure*}

Here we present the remaining comparisons between the true parameter values of \eagle\ and values recovered by the \magphys\ model when using the adjusted, older age prior distribution for \magphys\ seen in Fig.~\ref{fig: Age Hist}, for the 10,288 \eagle\ galaxies at redshift $z=0.10$. The comparisons for stellar mass and average stellar age are presented in Fig.~\ref{fig: OA Age and M*}. Here we show the comparisons for dust mass, dust luminosity, SFR, and sSFR in Fig.~\ref{fig: All Old Ages}. When compared to the recovered values of the redshift $z=0.10$ galaxies shown in Fig.~\ref{fig: Params 1} and Fig.~\ref{fig: Params 2}, the values of the median offsets, standard deviations and median error bars of these four parameters are very similar. This indicates that adjusting the age distribution of the \magphys\ age prior has little effect on the recovery of dust mass, dust luminosity, SFR and sSFR. There is a slight reduction in the median offset of sSFR, which is likely due to the reduction in the stellar mass offset when using the adjusted age prior.


\bsp	
\label{lastpage}
\end{document}